\documentclass[prc,preprint,
  showpacs,showkeys,lengthcheck,
  nofootinbib,tightenlines,twocolumn,notitlepage,
  preprintnumbers,superscriptaddress]{revtex4-1}

\usepackage[utf8]{inputenc}
\usepackage{newtxtext,newtxmath}
\usepackage[mathcal]{euscript}
\usepackage{ulem}

\usepackage{graphicx}
\usepackage[usenames,dvipsnames]{xcolor}
\graphicspath{{figures/}}

\usepackage{hyperref}
\hypersetup{colorlinks=true,citecolor=blue,linkcolor=blue,urlcolor=blue}

\usepackage{diagbox}
\usepackage{booktabs}

\usepackage{amsmath,amsfonts}
\usepackage{slashed}
\usepackage{bm}

\begin{document}

\title{Impact of DCSB and dynamical diquark correlations on proton GPDs}

\author{Adam Freese}
\email{afreese@anl.gov}
\address{Argonne National Laboratory, Lemont, Illinois 60439, USA}

\author{Ian C. Clo\"{e}t}
\email{icloet@anl.gov}
\address{Argonne National Laboratory, Lemont, Illinois 60439, USA}

\begin{abstract}
  We calculate the leading-twist, helicity-independent
  generalized parton distributions (GPDs) of the proton,
  at finite skewness,
  in the Nambu--Jona-Lasinio (NJL) model of quantum chomodynamics (QCD).
  The NJL model reproduces low-energy characteristics of QCD,
  including dynamical chiral symmetry breaking (DCSB).
  The proton bound-state amplitude is solved for using the Faddeev equation
  in a quark-diquark approximation,
  including both dynamical scalar and axial vector diquarks.
  GPDs are calculated using a dressed non-local correlator,
  consistent with DCSB,
  which is obtained by solving a Bethe-Salpeter equation.
  The model and approximations used observe Lorentz covariance,
  and as a consequence the GPDs obey polynomiality sum rules.
  The electromagnetic and gravitational form factors are
  obtained from the GPDs.
  We find a D-term of $-1.08$ when the non-local correlator is properly dressed,
  and $0.85$ when the bare correlator is used instead,
  suggesting that within this framework
  proton stability requires the constituent quarks to be dressed
  consistently with DCSB.
  We also find that the anomalous gravitomagnetic moment vanishes,
  as required by Poincar\'{e} symmetry.
\end{abstract}

\maketitle


\section{Introduction}

Generalized parton distributions (GPDs) appear in the calculation of hard
exclusive reactions such as deeply virtual Compton scattering (DVCS)
and deeply virtual meson production (DVMP).
Factorization \cite{Collins:1996fb,Collins:1998be,Ji:1998xh}
allows the amplitudes of these processes to be broken down
(up to power-suppressed corrections)
into the convolution of a hard scattering kernel and
a soft matrix element of quark and/or gluon fields
which contains the GPDs.

GPDs are Lorentz-invariant functions of four variables,
including the renormalization scale,
that describe many interesting properties of hadrons.
They encode spatial light cone distributions of partons through two-dimensional
Fourier transforms~\cite{Burkardt:2002hr}.
Additionally, their Mellin moments encode the electromagnetic and gravitational
properties of hadrons---allowing, in the latter case,
for such properties to be studied through hard exclusive reactions,
in lieu of the impossibility of graviton-exchange experiments.
The gravitational properties shed light on the way that
mass and angular momentum are distributed among the quarks and gluons
within the hadron, thus directly addressing deep questions about the
mass~\cite{Lorce:2018egm,Hatta:2018sqd}
and spin~\cite{Leader:2013jra}
decompositions of the proton.

In light of these considerations,
calculations of proton GPDs are to be highly desired.
It is vital that any model calculation observe the symmetries
and low-energy properties of quantum chromodynamics (QCD),
so that the qualitative and quantitative effects of these phenomena
manifest in the GPDs themselves.
For instance, the relationship between Mellin moments and
electromagnetic and gravitational form factors
is a consequence of Lorentz covariance~\cite{Ji:1996ek},
and the magnitude of the gravitational form factors has an intimate
relationship with dynamical chiral symmetry
breaking (DCSB)~\cite{Polyakov:2018zvc,Freese:2019bhb}.
For this reason, we use the Nambu--Jona-Lasinio (NJL) model of QCD
to perform calculations of proton GPDs.

The NJL model is an effective model of
quark interactions based on a four-fermi contact
interaction~\cite{Vogl:1991qt,Klevansky:1992qe,Hatsuda:1994pi}.
It successfully incorporates several low-energy aspects of QCD,
most notably (approximate) chiral symmetry and its dynamical breaking.
The breaking of chiral symmetry dresses the quarks,
causing them to propagate with a large effective mass $M\sim 400$~MeV,
as is described by the gap equation.
Moreover, confinement can be simulated in the NJL model through use of
proper time regularization~\cite{Ebert:1996vx,Hellstern:1997nv,Cloet:2014rja}.
The NJL model has been used to successfully describe many properties of both
mesons~\cite{Vogl:1991qt,Klevansky:1992qe,Cloet:2014rja,Ninomiya:2017ggn,Freese:2019bhb}
and baryons~\cite{Ishii:1993np,Ishii:1993rt,Ishii:1995bu,Cloet:2014rja}.

A particular aspect of proton structure we will emphasize is
the presence of diquark correlations.
Quark-diquark correlations have had considerable success in modeling
the properties of baryons~\cite{Cahill:1988dx,Anselmino:1992vg,Roberts:2011cf,Cloet:2014rja},
and the presence of diquark correlations is also borne out by such evidence as
the $Q^2$ dependence of flavor-separated form factors~\cite{Cates:2011pz}
and an approximate meson-baryon supersymmetry~\cite{Brodsky:2016rvj}.
We will thus calculate the proton's GPDs in a dynamical quark-diquark model.


\section{Proton GPDs in a dynamical quark-diquark model}

Generalized parton distributions (GPDs) are defined through the matrix elements
of bilocal light cone correlators.
The leading-twist, helicity-independent
quark GPDs of a hadron are defined through:
\begin{align}
  V_{\lambda\lambda'}^q
  &=
  \int \frac{\mathrm{d}\kappa}{2\pi}
  e^{2ix(Pn)\kappa}
  \left\langle p'\lambda' \middle|
  \overline\psi(-n\kappa)
  \slashed{n}
  [-n\kappa,n\kappa]
  \psi(n\kappa)
  \middle| p\lambda\right\rangle
  \,,
\end{align}
where $[x,y]$ is a Wilson line from $y$ to $x$,
$p$ and $p'$ are the initial and final momenta,
and
$\lambda$ and $\lambda'$ are the initial and final helicities (if applicable).
The GPDs themselves are obtained by decomposing $V_{\lambda\lambda'}^q$
in terms of linearly independent Lorentz structures.
For a spin-half hadron such as the proton, we have:
\begin{align}
  V_{\lambda\lambda'}^q
  =
  \bar{u}(p',\lambda')
  \left[
    \slashed{n}\, H(x,\xi,t)
    +
    \frac{i\sigma^{n\Delta}}{2M_N}\,
    E(x,\xi,t)
    \right]
  u(p,\lambda)
  \,,
\end{align}
where $P=\frac{p+p'}{2}$, $\Delta=p'-p$, $\xi = -2(\Delta n)/(Pn)$,
$t=\Delta^2$, and $n$ is a lightlike vector defining the light front.
The GPDs $H(x,\xi,t)$ and $E(x,\xi,t)$ are Lorentz-invariant functions
of the Lorentz-invariant arguments $x$, $\xi$, and $t$.
Similar correlators are defined for helicity-dependent and helicity-flip GPDs,
as well as for gluon GPDs.

\begin{figure}
  \centering
  \includegraphics[width=\columnwidth]{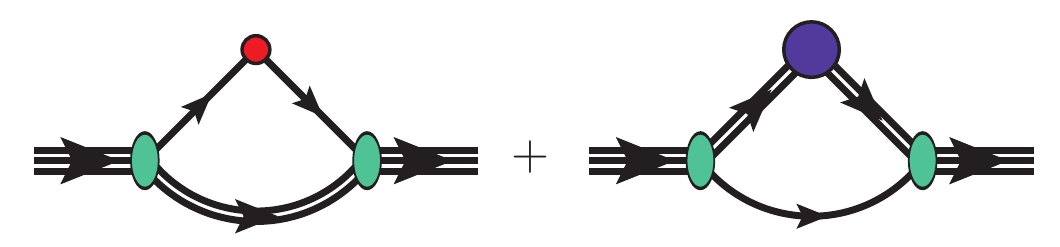}
  \caption{
    Diagrams contributing to the body GPDs of the proton
    within the static approximation for the interaction kernel.
    On the left is the direct quark diagram,
    and on the right is the diquark diagram.
    The (red) dot signifies an elementary vertex,
    and the (purple) blob a composite vertex.
  }
  \label{fig:proton:diagrams}
\end{figure}

The proton can be considered as a bound state of three constituent quarks,
which are dressed amalgamations of more elementary current quarks
(and, in QCD, gluons).
The bound state amplitude can be found by solving the Faddeev equation.
It has been found in many model
calculations~\cite{Cahill:1988dx}
that two of the three quarks are often bound in a diquark correlation
that is either isoscalar and Lorentz scalar, or isovector and Lorentz axial vector.
We will consider these configurations specifically in the calculations to follow.

A set of ``body GPDs'' (named in analogy to the ``body form factors''
in Ref.~\cite{Cloet:2014rja}) can be found by calculating the distribution
of these dressed quarks within the proton.
In momentum space, an operator of the form $\Gamma \delta(n[xP-k])$ 
is placed on each of the dressed quark lines,
with $\Gamma = \slashed{n}$ (times an isospin structure)
for the helicity-independent body GPDs.
Within the quark-diquark approximation,
the quark can be within or outside of the diquark correlation,
or possibly within the interaction kernel that binds the proton.
Within the approximations considered in this work,
the latter does not contribute to the GPDs.
The diagrams that do contribute are depicted in
Fig.~\ref{fig:proton:diagrams}.

Since GPDs describe the structure of hadrons in terms of current quarks,
the body GPDs are not by themselves sufficient.
Dressed quarks are not current quarks, but contain current quarks
as a more elementary substructure.
One can address this by dressing the bilocal operator defining the
light cone correlator, or, equivalently,
one can combine the body GPDs of the proton
with the GPDs of the dressed constituent quarks
using a convolution formula.
Such a convolution equation would also have applicability to describing
the non-elementary vertex in the diquark diagram of
Fig.~\ref{fig:proton:diagrams}.

Since it is of central importance to this work, we will first consider
how GPD convolution is to be done.
We will then explore the (body) GPDs and transition GPDs of the diquarks,
and subsequently the GPDs of the dressed constituent quarks.


\subsection{The convolution equation}
\label{sec:convolution}

\begin{figure}
  \centering
  \includegraphics[width=\columnwidth]{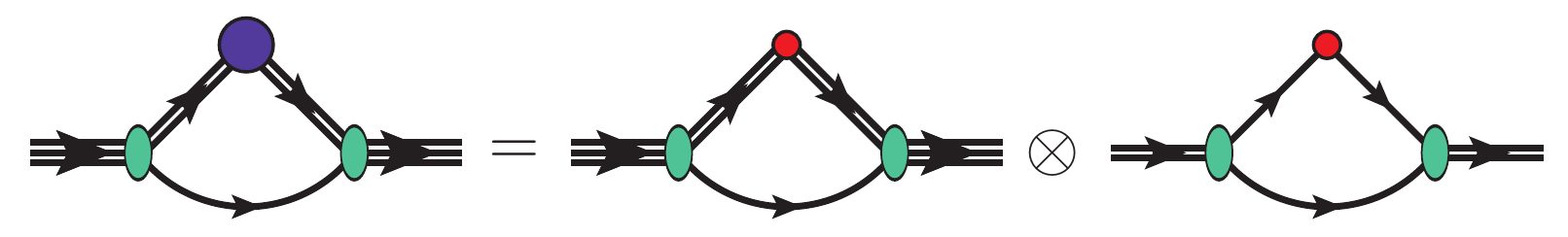}
  \caption{
    Diagrammatic depiction of the convolution equation.
    The large (purple) blob signifies a non-elementary GPD operator,
    while the small (red) dot signifies an elementary operator.
  }
  \label{fig:convolution}
\end{figure}

Let us consider a hadron $X$ to contain a composite constituent $Y$.
The insertion of a bilocal operator onto $Y$ can be expanded in
terms of the substructure of $Y$, as depicted in Fig.~\ref{fig:convolution}.
Since $Y$ is in general off its mass shell,
the composite operator (rightmost diagram in Fig.~\ref{fig:convolution})
is a function of the initial and final virtuality of $Y$.
In specific cases where there is no functional dependence on virtuality
(or when the dependence on virtuality can be safely neglected),
a convolution formula holds for the GPDs $H_{X,i}(x,\xi,t)$ of $X$:
\begin{align}
  H_{X,i}(x,\xi,t)
  =
  \sum_j
  \int \frac{\mathrm{d}y}{|y|}\,
  h_{Y/X,ij}(y,\xi,t)\,
  H_{Y,j}\left(\frac{x}{y},\frac{\xi}{y},t\right)
  \label{eqn:convolution}
  \,.
\end{align}
The indices $i$ and $j$ label the multiplicity of GPDs
appearing in front of the available Lorentz structures
for $X$ and $Y$, respectively.
$H_{Y,j}(x,\xi,t)$ are the GPDs of an on-shell $Y$.
The functions $h_{Y/X,ij}(y,\xi,t)$ are the body GPDs,
which are defined by using a Lorentz structure $\Gamma_j$ 
associated with the GPD $H_{Y,j}(x,\xi,t)$ in place of $\slashed{n}$
in the elementary bilocal operator.

When the constituents of $Y$ (which we call $Z$ for concreteness)
also have non-elementary substructure,
one can simply apply Eq.~(\ref{eqn:convolution}) consecutively.
It is possible to show that GPD convolution is associative, that is,
$
{\left( h_{Y/X} \otimes h_{Z/Y} \right) \otimes H_Z
=
h_{Y/X} \otimes \left( h_{Z/Y} \otimes H_Z \right)}
$,
so the consecutive applications of Eq.~(\ref{eqn:convolution})
can be done in either order.

In Sec.~\ref{sec:dressing}, we will observe that the dressed quark GPD has
no functional dependence on virtuality, allowing use of
Eq.~(\ref{eqn:convolution}) without caveats.
The GPD of an off-shell diquark does depend on virtuality in general,
but the use of a pole approximation for the diquark propagator requires
replacing vertices sandwiched between the propagators by their on-shell
form for consistency.


\subsection{Diquark body GPDs}

In order to determine the diquark diagram contribution to the proton body GPDs,
we must determine the body GPDs of the diquarks themselves.
We have both scalar and axial vector diquarks to consider,
in addition to the transition GPD between the two diquark species.

In this work, we use a pole approximation for the diquark propagators,
i.e., we take
\begin{align}
  \tau(p) \approx \tau_{\mathrm{pole}}(p)
  = \frac{1}{p^2 - M_{\mathrm{dq}}^2 + i0}
  \,,
\end{align}
which is the dominant contribution to the propagator.
Such approximations are ubiquitous in the baryon modeling
literature~\cite{Mineo:1999eq,Cloet:2005rt,Cloet:2005pp,Mineo:2005vs,Cloet:2006bq,Cloet:2007em,Eichmann:2007nn,Cloet:2008wg,Eichmann:2008ef,Nicmorus:2008vb,Nicmorus:2010sd,Matevosyan:2011vj,Roberts:2011cf,Wilson:2011aa,Cloet:2012cy,Cloet:2013gva,Segovia:2013uga,Segovia:2014aza,Segovia:2015hra,Carrillo-Serrano:2016igi}.
The operators $\Lambda(x,\xi,t,v,v')$ defining the diquark GPDs
in general depend on the initial and final virtualities:
\begin{align}
  v  = p^2  - M_{\mathrm{dq}}^2 \,, \qquad
  v' = p'^2 - M_{\mathrm{dq}}^2
  \,.
\end{align}
However, this operator is always found sandwiched between propagators in the form:
\begin{align}
  \tau(p') \Lambda(x,\xi,t,v,v') \tau(p)
  \label{eqn:sandwich}
  \,.
\end{align}
The pole approximation is in effect a truncated Laurent series expansion,
in $v$ (or $v'$) which is truncated above $v^{-1}$ ($v'^{-1}$).
When multiplying truncated series expansions,
the product must be truncated at the order of the approximation,
meaning that any order-$v$ (order-$v'$) terms in $\Lambda(x,\xi,t,v,v')$
must be discarded.
In other words, consistency with the pole approximation requires neglecting
the functional dependence of the GPD vertex on virtuality.
Thus, we replace the expression in Eq.~(\ref{eqn:sandwich}) with:
\begin{align}
  \tau(p') \Lambda(x,\xi,t,v,v') \tau(p)
  \approx
  \tau_{\mathrm{pole}}(p') \Lambda(x,\xi,t,0,0) \tau_{\mathrm{pole}}(p)
  \,,
\end{align}
and proceed to consider on-shell diquark GPDs.


\subsubsection{Scalar diquarks}

An on-shell scalar diquark has only a single GPD\footnote{
  In principle, an off-shell scalar diquark has
  an additional $\sf{T}$-odd GPD,
  but it is suppressed by the virtuality and, within the pole approximation,
  should be neglected.
}
which is identical for up and down quarks since the scalar diquark is isoscalar.
If one evaluates the far-right triangle diagram in Fig.~\ref{fig:convolution}
with a scalar Bethe-Salpeter vertex,
the matrix element decomposition is simply:
\begin{align}
  V_s^q
  =
  H_s^q(x,\xi,t)
  \,.
\end{align}
The corresponding body GPD for the distribution of scalar diquarks
within the proton is then found by evaluating the right diagram
in Fig.~\ref{fig:proton:diagrams}
with $(kn)\delta(n[xP-k])$ as the vertex.
This body GPD can be combined with $H_s^q(x,\xi,t)$
via Eq.~(\ref{eqn:convolution}) to obtain the
scalar diquark diagram contribution to the proton body GPDs.

The scalar diquark is isoscalar, so makes equal contributions to the
up and down body GPDs.
Additionally, since it contains the proton's valence down quark,
diagrams with a spectator scalar diquark only contribute to the up
quark body GPD (which can still contain a down current quark).


\subsubsection{Axial vector diquarks}

The axial vector diquark has five on-shell GPDs.
The relevant Lorentz decomposition is \cite{Berger:2001zb}:
\begin{multline}
  V_{a,\lambda^\prime\lambda}^q
  =
  -(\epsilon\epsilon'^*)
  H_{1a}^q
  + \frac{
    (\epsilon'^* n)(\epsilon\Delta)
    -
    (\epsilon n)(\epsilon'^*\Delta)
  }{2(Pn)}
  H_{2a}^q
  \\
  + \frac{(\epsilon\Delta)(\epsilon'^*\Delta)}{2M^2}
  H_{3a}^q
  - \frac{
    (\epsilon'^* n)(\epsilon\Delta)
    +
    (\epsilon n)(\epsilon'^*\Delta)
  }{2(Pn)}
  H_{4a}^q
  \\
  + \left[ \frac{M^2(\epsilon n)(\epsilon'^* n)}{(Pn)^2}
  + \frac{1}{3}(\epsilon\epsilon'^*)\right]
  H_{5a}^q
  \,,
\end{multline}
where the functional dependence on $x$, $\xi$, and $t$ has been suppressed
for compactness.

One can obtain an off-shell version of this by not including the polarization
vectors in the calculation of $V_{a,\lambda^\prime\lambda}^q$.
There appears to be an ambiguity in this,
since prior to stripping the polarization vectors,
the identities $(\epsilon p)=(\epsilon'^* p')=0$
can be used to rewrite (for instance)
$(\epsilon\Delta)$ in terms of $(\epsilon P)$.
However, within the pole approximation,
the axial vector diquark propagators are transverse,
thus enforcing the similarity relations $p^\mu\sim0$ and $p'^\nu\sim0$.
We may use these similarity relations to rewrite the uncontracted correlator
$V_{a,\lambda^\prime\lambda}^{q,\mu\nu}$
with $\Delta$ as the only uncontracted momentum,
giving us:
\begin{multline}
  V_{a,\lambda^\prime\lambda}^{q,\mu\nu}
  =
  -g^{\mu\nu}
  H_{1a}^q
  + \frac{
    n^\nu \Delta^\mu
    -
    n^\mu \Delta^\nu
  }{2(Pn)}
  H_{2a}^q
  + \frac{\Delta^\mu\Delta^\nu}{2M^2}
  H_{3a}^q
  \\
  - \frac{
    n^\nu \Delta^\mu
    +
    n^\mu \Delta^\nu
  }{2(Pn)}
  H_{4a}^q
  + \left[ \frac{M^2 n^\mu n^\nu}{(Pn)^2}
  + \frac{1}{3}g^{\mu\nu} \right]
  H_{5a}^q
  \label{eqn:gpd:axial:offshell}
  \,.
\end{multline}
The Lorentz structures above can be used to calculate the body GPDs
for the distribution of axial vector diquarks within the proton,
provided the substitution $P\mapsto k$ is made,
and the structures are then multiplied by $(kn)\delta(n[xP-k])$.

The axial vector diquark is isovector,
meaning it comes in $uu$, $ud$, and $dd$ varieties.
The isospin algebra necessary to determine the weights with which
diagrams involving axial vector diquarks enter into up and down quark body GPDs
has previously been done,
with the results in Eqs.~(102,103) of Ref.~\cite{Cloet:2014rja}.\footnote{
  Although Ref.~\cite{Cloet:2014rja} is about form factors,
  the isospin algebra is the same, and the relevant isospin factors
  are also the same.
}


\subsubsection{Diquark transition GPDs}

Lastly, there are scalar-to-axial and an axial-to-scalar transition GPDs.
For scalar $\rightarrow$ axial vector transitions, the correlator takes the form:
\begin{align}
  V_{sa}^q
  =
  \frac{1}{(Pn)}
  \frac{1}{M_s+M_a}
  i\epsilon^{P\Delta n \epsilon}
  H_{sa}^q(x,\xi,t)
  \,.
\end{align}
Hermiticity and time reversal properties can be used to show that
$H_{sa}^q(x,\xi,t) = H_{sa}^q(x,-\xi,t)$
and that the axial vector $\rightarrow$ scalar transition GPD satisfies
$H_{as}^q(x,\xi,t) = -H_{sa}^q(x,\xi,t)$.
Neglecting virtuality dependence (as required by the pole approximation),
there remains one GPD in the off-shell case,
since stripping the polarization vector $\epsilon$ from $V_{sa}^q$ can still
only produce a single unique Lorentz structure.

As for the axial vector diquark, the isospin weights for the transition diagrams
can be found in Eqs.~(102,103) of Ref.~\cite{Cloet:2014rja}.


\subsection{Dressed quark GPDs}
\label{sec:dressing}

In order to calculate any hadronic matrix element within the NJL model,
one must dress the operator in question.
This dressing is a result of DCSB,
and is just as necessary for bilocal light cone correlators and GPDs
as it is for the electromagnetic current and form factors.
Failing to dress the operator defining GPDs will result in its Mellin
moments reproducing the incorrect electromagnetic and gravitational form
factors~\cite{Freese:2019bhb,shi:inpress}.
The need for dressing arises because 
the GPDs appear in a bilocal correlator of current quark fields,
while a hadron in the NJL model is constructed from composite dressed quarks.
Dressing of the operator essentially amounts to describing the
dressed quark in terms of an elementary current quark substructure.

\begin{figure}
  \centering
  \includegraphics[width=\columnwidth]{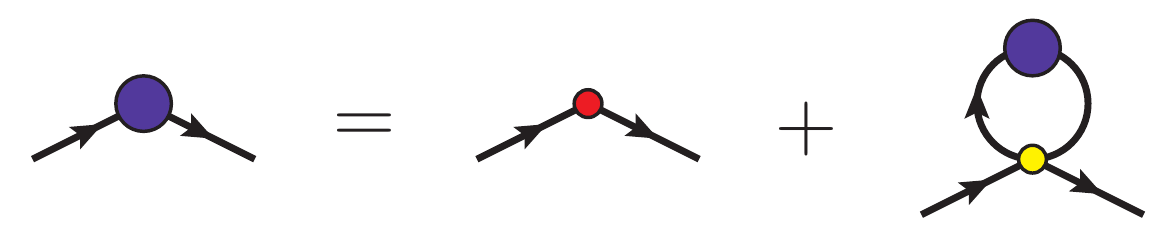}
  \caption{
    Diagrammatic depiction of the Bethe-Salpeter equation for the
    bilocal light cone operator defining leading-twist GPDs.
  }
  \label{fig:bse}
\end{figure}

The bilocal operator defining the leading-twist GPDs is dressed according
to a Bethe-Salpeter equation (BSE),
which is depicted in the Hartree-Fock approximation\footnote{
  This approximation is standard for NJL model
  calculations~\cite{Vogl:1991qt,Klevansky:1992qe,Hatsuda:1994pi},
  and excludes diagrams with more than one loop.
} by Fig.~\ref{fig:bse}.
One can see this by considering Mellin moments of the bilocal operator,
which define local operators that each obey a BSE
that is also depicted by Fig.~\ref{fig:bse}.
At leading twist, these local operators are traceless and accordingly
do not receive contributions from operators inserted on the four-point vertex.
Due to the uniqueness of the inverse Mellin transform,
the bilocal operator is likewise dressed according to Fig.~\ref{fig:bse}.

This BSE holds for both up and down quarks,
and the NJL interaction kernel mixes these equations.
It is most straightforward to solve the decoupled BSEs for the isoscalar
and isovector GPDs $H_I(x,\xi,t)$, defined as:
\begin{subequations}
  \begin{align}
    H_{I=0}(x,\xi,t) &= H_u(x,\xi,t) + H_d(x,\xi,t) \,, \\
    H_{I=1}(x,\xi,t) &= H_u(x,\xi,t) - H_d(x,\xi,t)
    \,,
  \end{align}
\end{subequations}
and analogously for $E_{I}(x,\xi,t)$,
which appear in Lorentz decompositions of the operators
$\slashed{n}\delta(n[xP-k])$ and $\slashed{n}\tau_3\delta(n[xP-k])$,
respectively.

We find the following solutions for the dressed quark GPDs:
\begin{subequations}
  \begin{align}
    H_{I}(x,\xi,t)
    &=
    \delta(1-x)
    +
    H'_{I}(x,\xi,t)
    +
    \delta_{I,0} D_Q(x,\xi,t)
    \,,
    \\
    E_{I}(x,\xi,t)
    &=
    -\delta_{I,0} D_Q(x,\xi,t)
    \,,
    \\
    H^\prime_{I=0,1}(x,\xi,t)
    &=
    \frac{N_c}{4\pi^2}
    \frac{1}{|\xi|}
    \frac{
      G_{\omega,\rho} t\left(1-x^2/\xi^2\right)
    }{1 + 2G_{\omega,\rho} \Pi_{VV}(t)}
    \mathcal{E}_1\left(\frac{x}{\xi},t\right)
    \Theta(|\xi|-|x|)
    \label{eqn:gpd:HQ}
    \,,
    \\
    D_Q(x,\xi,t)
    &=
    -
    \frac{N_c}{\pi^2}
    \frac{x}{|\xi|}
    \frac{G_\pi M^2 }{ 1 - 2G_\pi \Pi_{SS}(t) }
    \mathcal{E}_1\left(\frac{x}{\xi},t\right)
    \Theta(|\xi|-|x|)
    \,,
    \label{eqn:gpd:DQ}
  \end{align}
  \label{eqn:gpd:quark}
  where
  \begin{align}
    \mathcal{E}_1\left(z,t\right)
    =
    E_1\left(\frac{4M^2-t(1-z^2)}{4\Lambda_{\mathrm{UV}}^2}\right)
    -
    E_1\left(\frac{4M^2-t(1-z^2)}{4\Lambda_{\mathrm{IR}}^2}\right)
    \,,
  \end{align}
\end{subequations}
and $E_1(z) = \int_1^\infty\mathrm{d}t\,t^{-1}e^{-zt}$
is the exponential integral function.
These solutions are exact, and do not contain any explicit functional dependence
on the quark virtuality, despite the quark being off-shell in general.
This is a consequence of the interaction that produces the dressing being a
contact interaction.

It is worth remarking on the limit $\xi\rightarrow0$ in Eq.~(\ref{eqn:gpd:quark}).
It can be shown that the integral of $H'_I(x,\xi,t)$ over $x$ is independent
of $\xi$ when $\xi>0$, and that all higher Mellin moments contain an overall
factor $\xi$, and thus vanish when $\xi\rightarrow0$.
From the uniqueness of the Mellin transform, we infer that when $\xi\rightarrow0$,
$H_I'(x,\xi,t)$ is proportional to a Dirac delta distribution.
This was also found in Ref.~\cite{shi:inpress},
where it was remarked that even in the zero-skewness limit,
the GPD contains a ``hidden ERBL region'' at $x=0$.
This hidden ERBL region persists through GPD convolution,
meaning that numerical results at $\xi=0$ for hadron GPDs in the NJL model
will necessarily be missing the delta distribution in the hidden ERBL region.


\subsection{Support region}

The support region in $x$ of the GPDs and body GPDs
is determined during the course of evaluating the relevant diagrams.
Each diagram's contribution is non-zero only when the delta
function $\delta(n[xP-k])$ is picked up by the integration over $k$.
We find in particular that:
\begin{align}
  -|\xi| \leq x \leq \max(1,|\xi|)
  \label{eqn:support}
  \,,
\end{align}
except in the case of the GPDs of dressed quarks,
for which $|x| < |\xi|$, as is explicitly noted by the presence of
step functions in Eqs.~(\ref{eqn:gpd:quark}).
The condition $|\xi| \leq 1$ holds for any on-shell particle
by virtue of kinematics,
which would strengthen Eq.~(\ref{eqn:support}) to $x \in [-|\xi|,1]$.
However, it is possible to have $|\xi| > 1$ for off-shell particles.
To see this, consider that:
\begin{align}
  \xi = \frac{(np)-(np')}{(np)+(np')}
  \,.
\end{align}
For an on-shell particle, $(np) \propto E_p + p_z$ is strictly non-negative,
so the constraint $|\xi| \leq 1$ follows from the triangle inequality.
On the other hand, off-shell particles are not required to satisfy any
such constraint, and in fact $(np)$ can be negative.
Thus $\xi$ is not constrained in general.

Since we consider off-shell particles in using the convolution formula
Eq.~(\ref{eqn:convolution}),
we leave the support region in Eq.~(\ref{eqn:support}) as general as possible.
In particular, since $y \sim 0$ is present in the integral,
the off-shell constituent $Y$ can have arbitrarily large skewness.
We also find in numerical calculations using Eq.~(\ref{eqn:convolution})
that having support at $x>1$ for $H_Y(x,\xi,t)$ is necessary for $H_X(x,\xi,t)$
to satisfy polynomiality (see Sec.~\ref{sec:poly}).


\section{Proton GPD results}
\label{sec:results}

Several variations of the NJL model exist.
In Ref.~\cite{Cloet:2014rja}, the electromagnetic properties of the proton
were found to be well-described within a two-flavor variant of the model.
We thus use the model variant described in Ref.~\cite{Cloet:2014rja},
including the numerical values for the model parameters and
the approximations described therein,
to calculate the helicity-independent, leading-twist proton GPDs.

\begin{figure}
  \centering
  \includegraphics[width=\columnwidth]{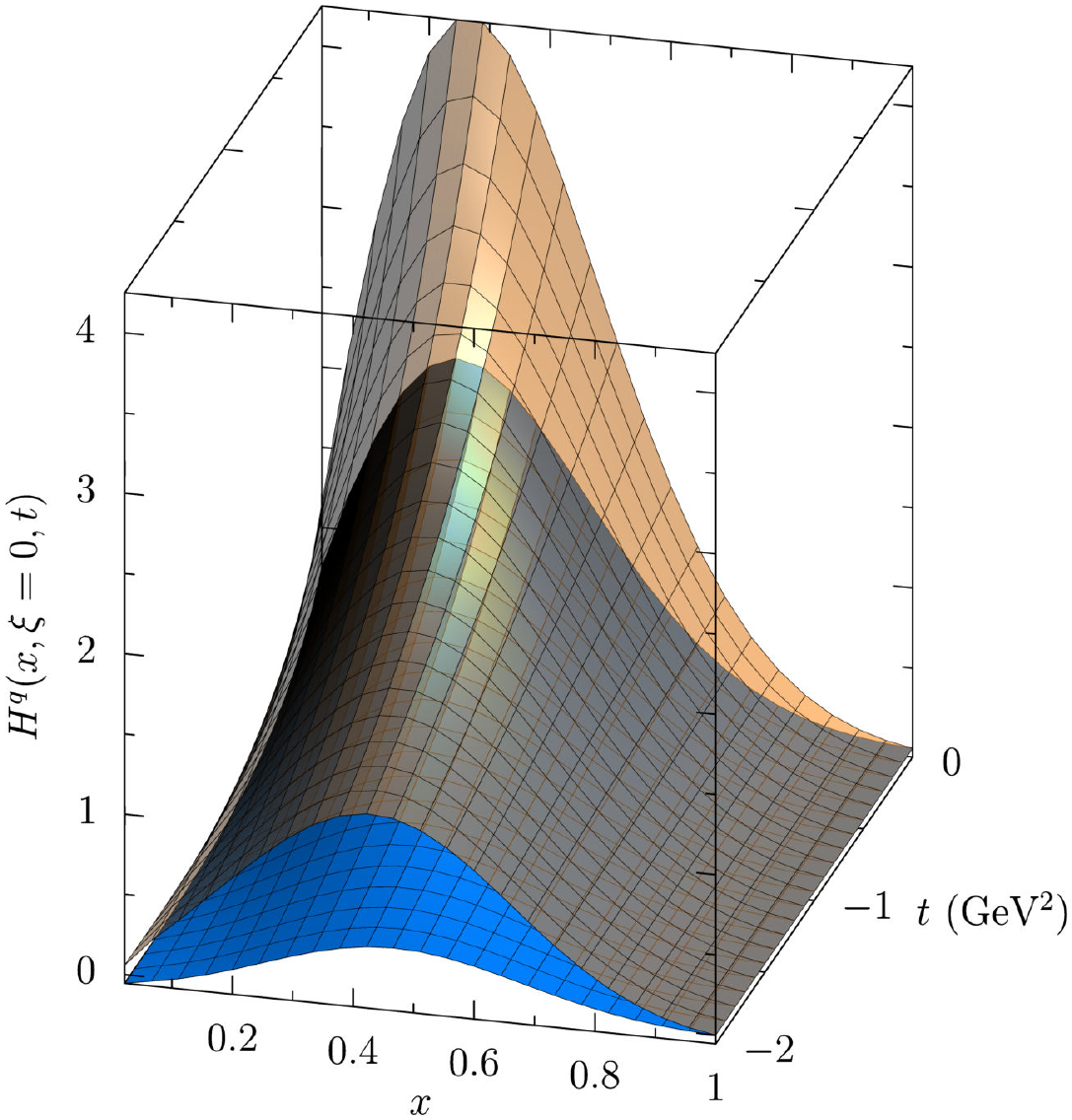}
  \includegraphics[width=\columnwidth]{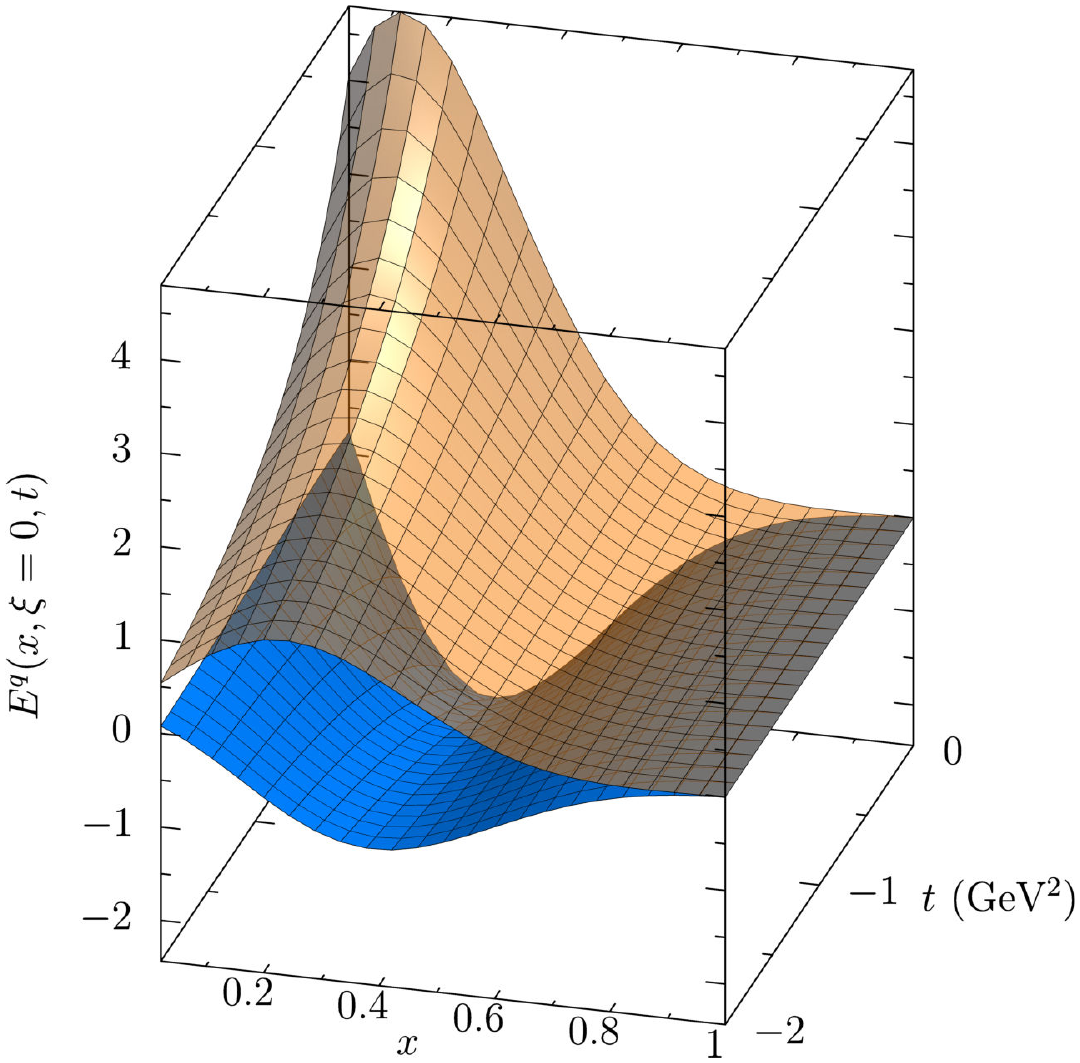}
  \caption{
    Proton GPD results at the model scale ($\mu^2=0.16$~GeV$^2$),
    and $\xi=0$, as a function of $x$ and $t$.
    Transparent (orange) surfaces are up quark distributions,
    opaque (blue) are down quark.
  }
  \label{fig:gpds:noskew}
\end{figure}
\begin{figure}
  \centering
  \includegraphics[width=\columnwidth]{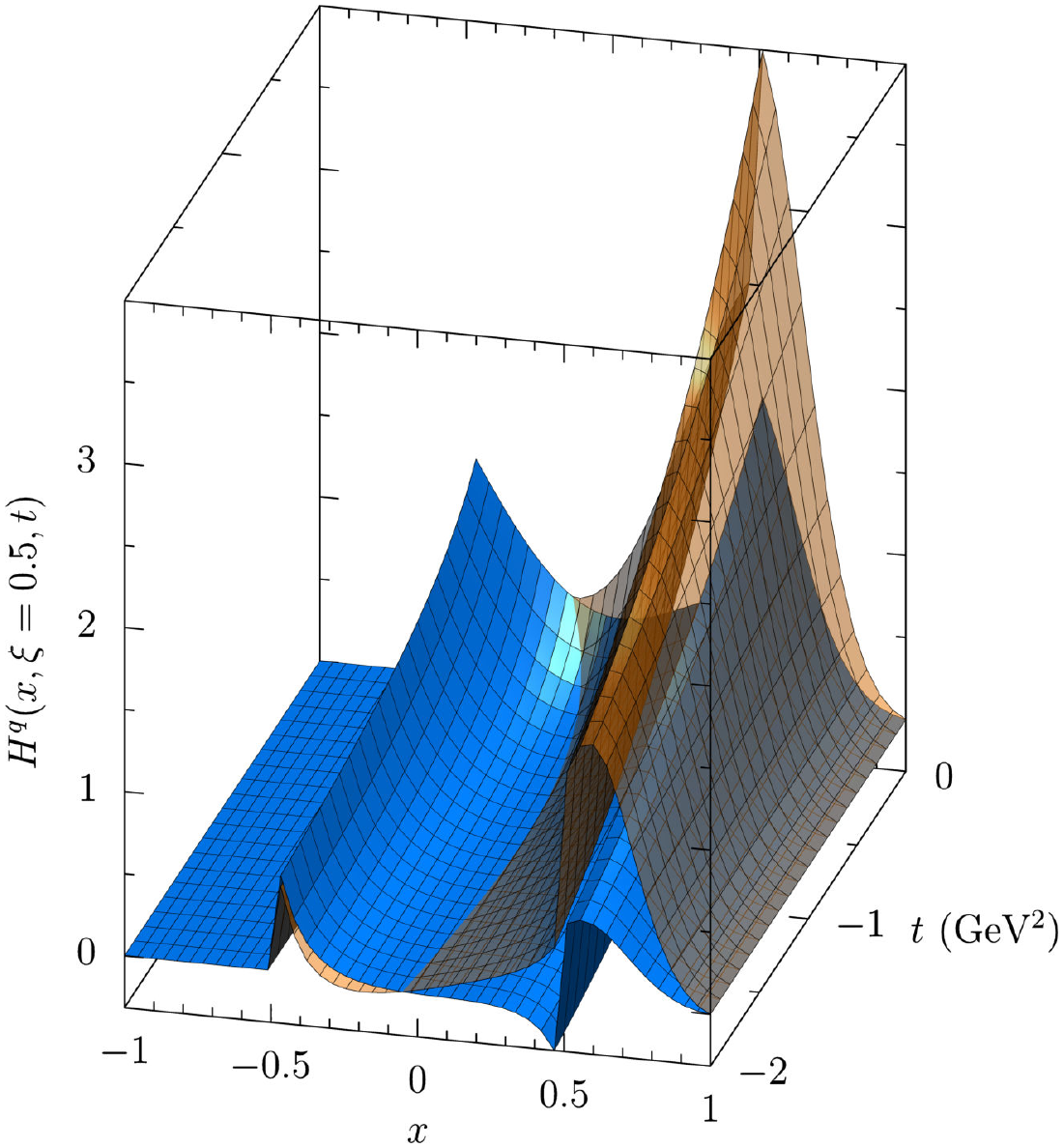}
  \includegraphics[width=\columnwidth]{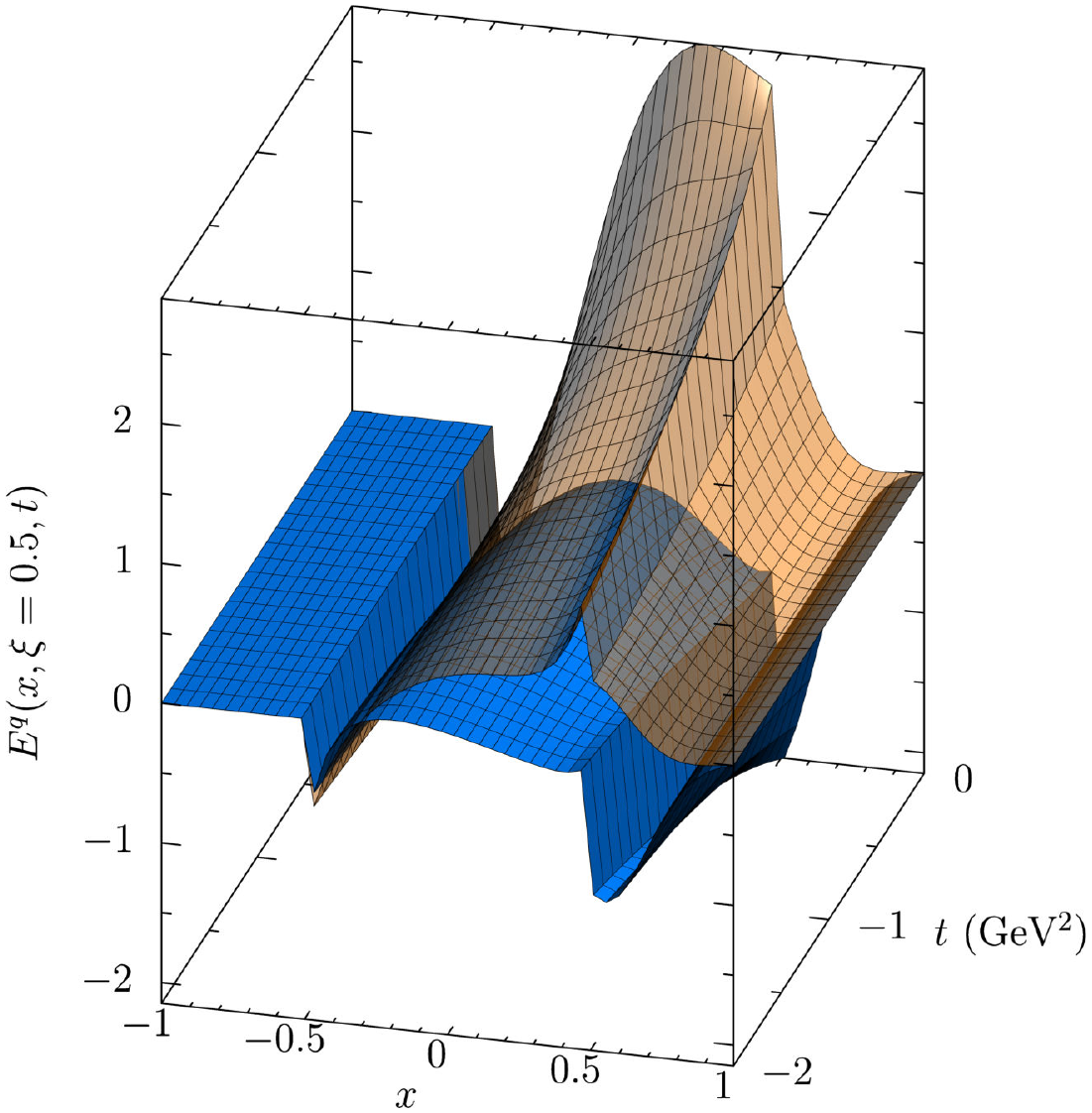}
  \caption{
    Proton GPD results at the model scale ($\mu^2=0.16$~GeV$^2$),
    and $\xi=0.5$, as a function of $x$ and $t$.
    Surfaces have the same meaning as in Fig.~\ref{fig:gpds:noskew}.
  }
  \label{fig:gpds:skew}
\end{figure}

We begin by presenting results for the GPDs $H^q(x,\xi,t)$ and $E^q(x,\xi,t)$ 
at two skewness values,
and at a model renormalization scale of $\mu^2=M^2=0.16$~GeV$^2$.
In Fig.~\ref{fig:gpds:noskew}, we have $\xi=0$,
while in Fig.~\ref{fig:gpds:skew}, we consider a moderate $\xi=0.5$.

To help understand the results, we first give the relationships
that the GPDs have to more familiar observables.
For instance, we have the forward limit relation $H_q(x,0,0) = q(x)$.
Additionally, Mellin moments of the GPDs are related to
the electromagnetic form factors (EMFFs):
\begin{align}
  \label{eqn:emffs}
  \int \mathrm{d}x\, H^q(x,\xi,t) = F^q_1(t), \qquad
  \int \mathrm{d}x\, E^q(x,\xi,t) = F^q_2(t),
\end{align}
where $F_i(t) = \sum_q e_q F_i^q(t)$,
and to the gravitational form factors (GFFs):
\begin{subequations}
  \label{eqn:gffs}
  \begin{align}
    \int \mathrm{d}x\,
    x
    H^{q}(x,\xi,t)
    &=
    A^q(t) + \xi^2 C^q(t)
    \,,
    \\
    \int \mathrm{d}x\,
    x
    E^{q}(x,\xi,t)
    &=
    B^q(t) - \xi^2 C^q(t)
    \,.
  \end{align}
\end{subequations}
An angular momentum form factor
$J^q(t) = \frac{1}{2}\left(A^q(t) + B^q(t)\right)$ can also be defined,
and is related to Ji's sum rule~\cite{Ji:1996ek}.

The $x$ and $t$ dependence of the GPDs in both
Figs.~\ref{fig:gpds:noskew},\ref{fig:gpds:skew} can be seen to differ
between the up and down quarks.
This is due to the presence of multiple isospin-dependent effects.
Among these is the presence of diquark correlations,
whose effects can be most easily seen in the non-skewed GPDs.

At zero skewness ($\xi=0$), there is a peak in $H^q(x,0,t)$
for each fixed-$t$ slice.
(See top panel of Fig.~\ref{fig:gpds:noskew}.)
The location for this peak can be quantified by the average momentum fraction
$\langle\langle x_q(t) \rangle\rangle = A^q(t) / F_1^q(t)$.
In the forward limit,
$\langle\langle x_u(0)\rangle\rangle = 0.34
\approx
\langle\langle x_d(0)\rangle\rangle = 0.32$.
Each down quark thus carries about the same momentum on average than each up quark.
Were only scalar diquarks present,
we would expect $\langle\langle x_u(0)\rangle\rangle \gg \langle\langle x_d(0)\rangle\rangle$,
since a (dressed) down quark would only be found within the diquark,
thus giving the down quark a lower effective mass.
However, axial vector diquark configurations are also present, and
(due to how the recoupling coefficients work out)
the down quark is more often alone than in the axial vector diquark.
Thus, the difference between
$\langle\langle x_u(0)\rangle\rangle$
and $\langle\langle x_d(0)\rangle\rangle$
is softened by the presence of axial vector diquarks.

Finite $t$ is a novel aspect of GPDs that elaborates the roles of
different diquark species further.
High $-t$ acts as a filter that selects for configurations where the probed parton
was already carrying most of the hadron's momentum.
One can accordingly observe in Fig.~\ref{fig:gpds:noskew}
that increasing $-t$ moves the peaks of both GPDs to higher $x$.
The down quark peak migrates further than the up quark,
with $\langle\langle x_u(-2~\mathrm{GeV}^2)\rangle\rangle = 0.45$
and $\langle\langle x_d(-2~\mathrm{GeV}^2)\rangle\rangle = 0.50$.
This occurs because axial vector diquark configurations fall more slowly with $-t$,
so at large $-t$ the down quark becomes sampled more often outside a diquark.

Despite $\langle\langle x_d(t)\rangle\rangle$
exceeding $\langle\langle x_u(t)\rangle\rangle$
at large $-t$, the down quark GPD still falls faster than the up quark GPD
with increasing $-t$,
as has previously been seen in
measurements of the flavor-separated electromagnetic
form factors~\cite{Cates:2011pz}.
We define the ratio
$
  S_q(x,t) = H_q(x,0,t) / H_q(x,0,0)
$,
which characterizes how quickly an $x$ slice of a quark GPD falls with $-t$.
The super-ratio $S_d(x,t) / S_u(x,t)$ then characterizes how much faster
the down quark GPD falls with $-t$ than the up quark GPD.

\begin{figure}
  \centering
  \includegraphics[width=\columnwidth]{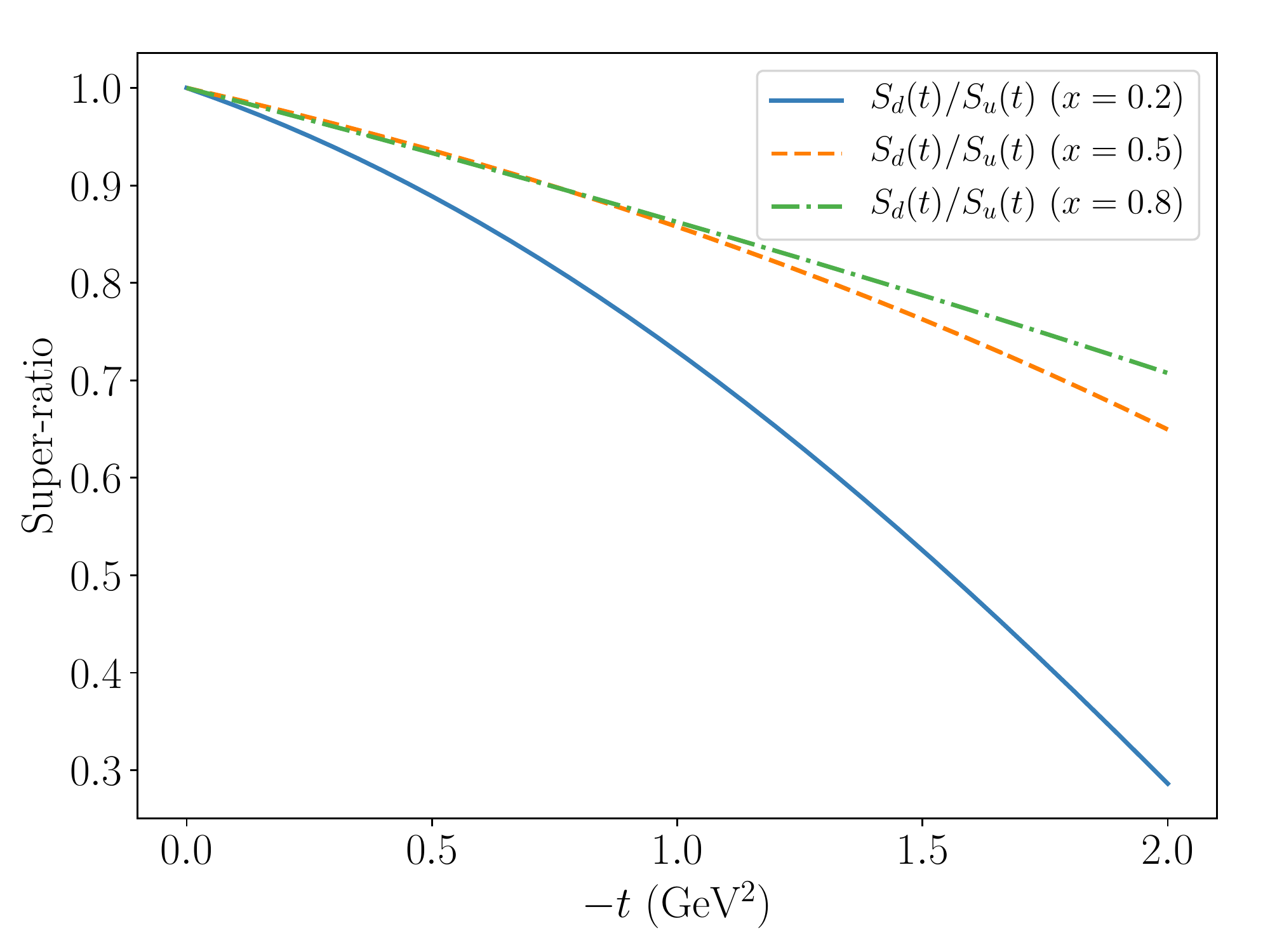}
  \caption{
    The super-ratio $S_d(x,t) / S_u(x,t)$ for several values of $x$.
  }
  \label{fig:super-ratio}
\end{figure}

Such a super-ratio is plotted in Fig.~\ref{fig:super-ratio} for several
values of $x$.
We find at all $x$ values that the down quark GPD falls more steeply
than the up quark GPD,
but that the steepness itself varies as a function of $x$.
In particular, the relative steepness of the down quark falloff
is less extreme at large $x$ than at small $x$.
Since the steeper down quark falloff is a consequence of diquark correlations,
this suggests that diquark correlations dominate at small $x$,
and one is more likely to observe a direct quark outside the diquark correlation
when $x$ is close to 1.

The other GPD, $E^q(x,\xi,t)$, does not correspond to a familiar observable
in the forward limit, but can be understood by its relationships to the
form factors $F_2^q(t)$, $B^q(t)$, and $J^q(t)$.
Unlike with $H^q(x,\xi,t)$, the peaks for $E^q(x,\xi,t)$ are at lower
$x$ for the up quarks than down quarks for all values of $t$.
(This can be clearly seen in the lower panel of Fig.~\ref{fig:gpds:noskew}.)
This occurs due to axial vector diquarks having a larger impact on
$E^q(x,\xi,t)$ than scalar diquarks.

Another novel aspect of GPDs is finite skewness.
We emphasize that a fully covariant calculation is necessary for a correct
description at finite skewness,
particularly in the ERBL region, where $|x| < |\xi|$.
A popular non-covariant method to calculate GPDs is the overlap representation
using a truncated light front basis expansion~\cite{Diehl:2000xz}.
The truncation in particular is not invariant under the full
Lorentz group~\cite{deMelo:1998an,Li:2017uug},
and moreover leads to missing terms in the overlap representation of the ERBL
region,
since the GPD in the ERBL region is generated by the overlap of Fock states
with different numbers of partons.

\begin{figure}
  \centering
  \includegraphics[width=\columnwidth]{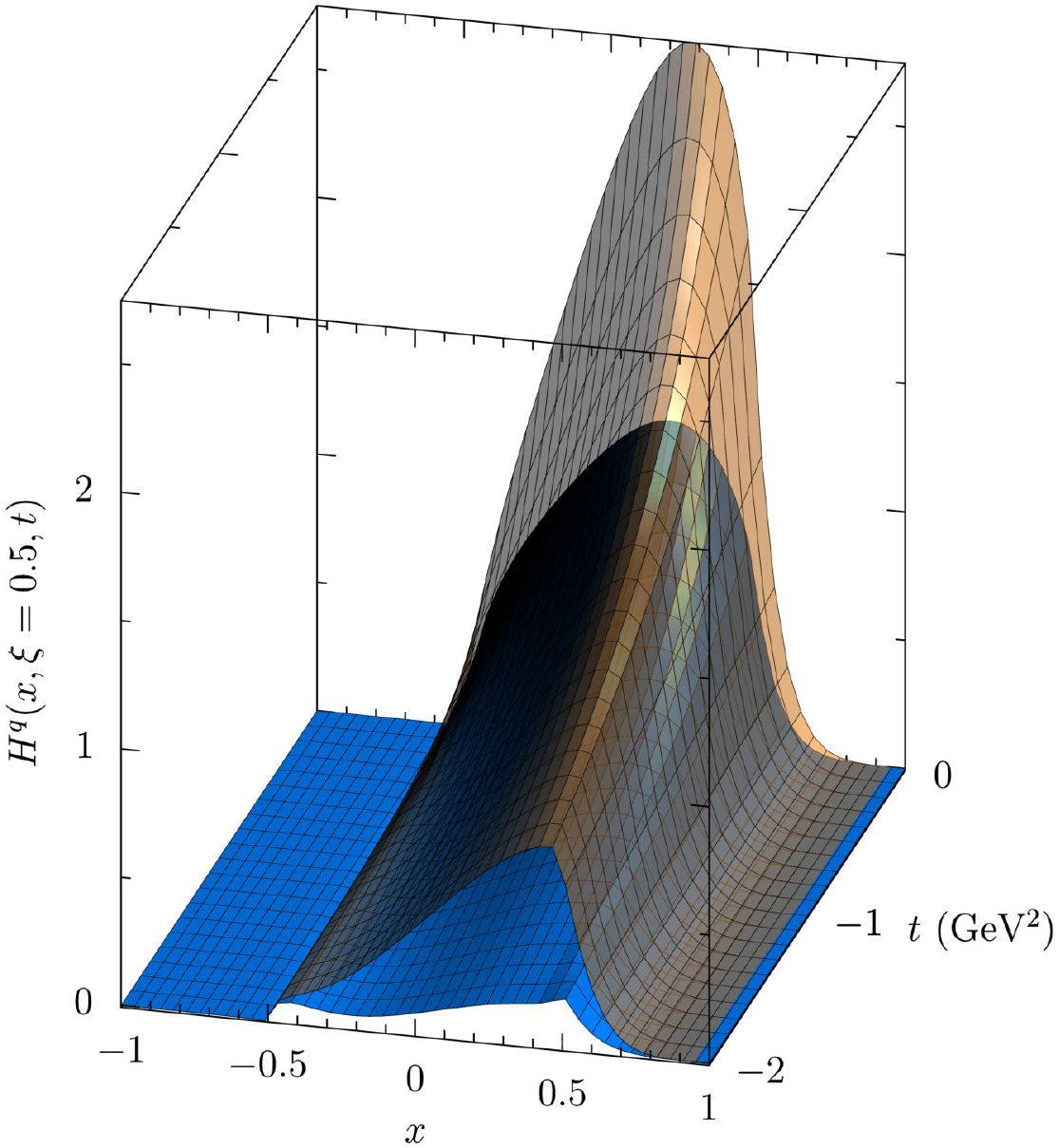}
  \includegraphics[width=\columnwidth]{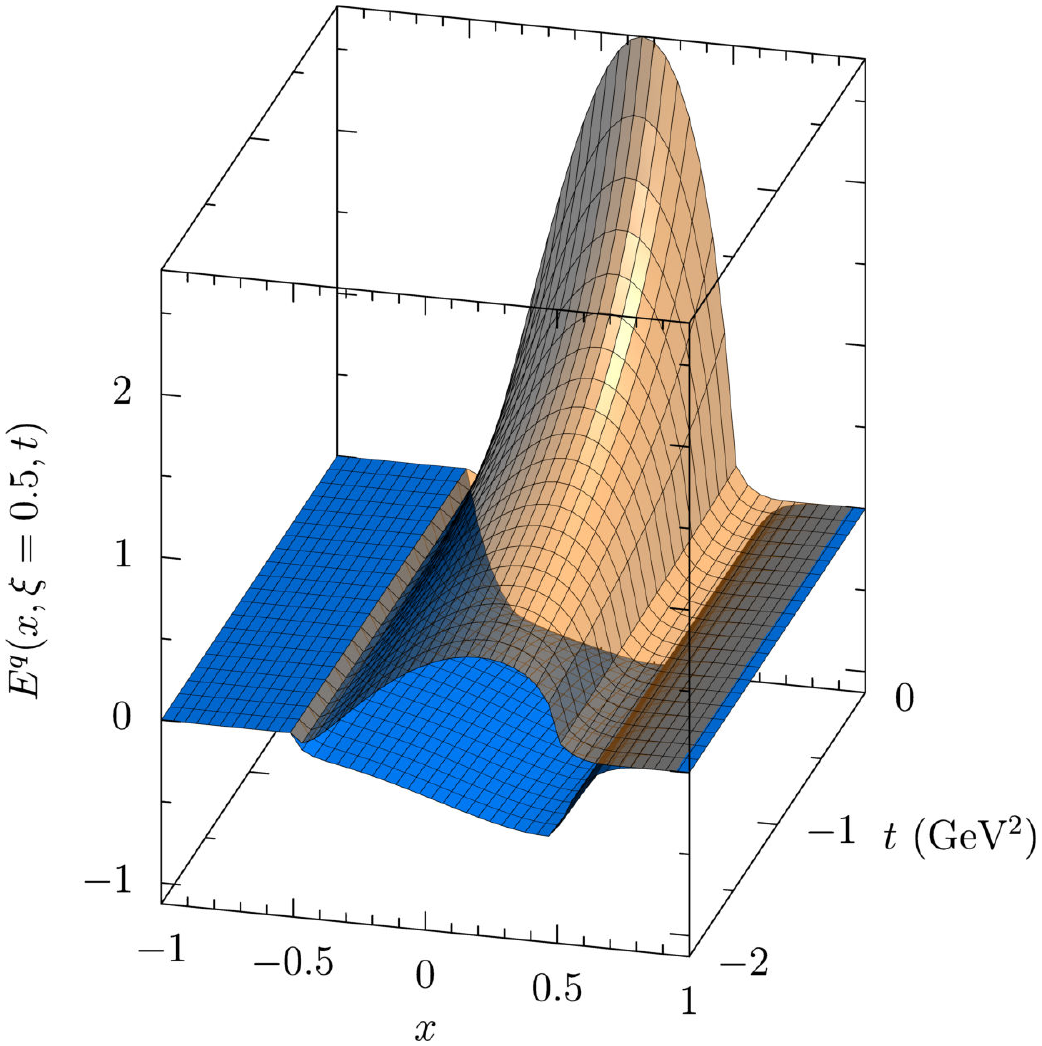}
  \caption{
    Proton GPD results at the an evolved scale ($\mu^2=4$~GeV$^2$),
    and $\xi=0.5$, as a function of $x$ and $t$.
    Surfaces have the same meaning as in Fig.~\ref{fig:gpds:noskew}.
  }
  \label{fig:gpds:evolved}
\end{figure}

Since the calculations done in this work are fully covariant,
we are able to obtain self-consistent results at finite skewness,
including in the ERBL region.
In Fig.~\ref{fig:gpds:skew}, GPD results are shown for $\xi=0.5$.
The most immediately striking feature are the jump discontinuities at $x=\pm\xi$.
These discontinuities are inherited from the dressed quark GPDs
given in Eq.~(\ref{eqn:gpd:quark}),
and are not present if the quark GPDs are not dressed.
These jump discontinuities are characteristic of effective model calculations
with constant dressed quark mass,
and have previously been observed in other model GPD
calculations~\cite{Petrov:1998kf,Polyakov:1999gs,Theussl:2002xp}.
These discontinuities appear to present a problem for factorization
of the DVCS amplitude, but factorization is not expected
to apply at the model scale of $\mu^2=0.16$~GeV$^2$
and evolution of the model GPDs to a scale where QCD factorization
is expected to apply removes the discontinuities
(as we will see in Fig.~\ref{fig:gpds:evolved}).

Another striking feature of Fig.~\ref{fig:gpds:skew} is the 
drastic difference in the $x$ dependence of the up and down GPDs
in the ERBL region, where $|x| < |\xi|$.
In this region, the dressed quark GPD is no longer trivial,
and accordingly it is possible to find (for instance) a down current quark
inside a dressed up quark.

At large $-t$, the down quark GPD $H_d(x,\xi,t)$ is dominated by
$D_Q(x,\xi,t)$ in the ERBL region.
This happens because the down quark body GPD falls faster,
due to the dominance of scalar diquark configurations,
causing $H_d(x,\xi,t)$ to be dominated by the current down quarks
found within dressed up quarks.
Moreover, the GPD for a current quark to appear within a
dressed quark of the opposite flavor goes as
$\frac{1}{2}( H_{I=0}(x,\xi,t) - H_{I=1}(x,\xi,t) + D_Q(x,\xi,t) )$,
where $H_{I=0}(x,\xi,t) \approx H_{I=1}(x,\xi,t)$ because of the nearly
degenerate masses of the $\rho$ and $\omega$ mesons.
Thus, $D_Q(x,\xi,t)$, which is an odd function of $x$ and is negative
for $x > 0$, dominates the ERBL region of $H_d(x,\xi,t)$ at large $-t$.

Many of the peculiar model features are washed out by GPD evolution.
In Fig.~\ref{fig:gpds:evolved}, we present results of evolving the
model scale GPDs to $\mu^2 = 4$~GeV$^2$.
Leading-order evolution kernels~\cite{Ji:1996nm} were used
along with a zero-mass variable flavor number scheme.
The jump discontinuities at $x=\pm\xi$ are removed,
and the shape of $D_Q(x,\xi,t)$ is no longer clearly visible in the
ERBL region.
Since the jump discontinuities disappear at $Q^2$ above the model scale,
the model scale discontinuities do not present a problem for factorization
at scales where QCD factorization is expected to be relevant.

\subsection{Polynomiality and form factors}

\label{sec:poly}

An especially remarkable property of GPDs is polynomiality,
which ensures that the $s$th Mellin moment of a GPD is a polynomial
in $\xi$ of order $s$ or less\footnote{
  This is true for the gluon GPD if the Ji convention~\cite{Ji:1998pc}
  is used. If the Diehl convention~\cite{Diehl:2003ny} is used,
  this is true instead of the $(s-1)$th Mellin moment.
}.
For the helicity-independent GPDs of spin-half particles,
the polynomials are even in $\xi$\footnote{
  GPDs that are odd in $\xi$ exist, even at leading twist.
  $H_4(x,\xi,t)$ and $\tilde{H}_3(x,\xi,t)$ for spin-one targets are
  such examples~\cite{Diehl:2003ny}.
  The matrix element of the bilocal correlator must be time reversal invariant,
  and if it's possible to construct T-odd Lorentz structures in its decomposition,
  the Lorentz-invariant function multiplying it must likewise be T-odd
  (i.e., odd in $\xi$) for the product of both to be T-even.
}.
Polynomiality is a consequence of Lorentz covariance,
with the reality of the coefficients following from hermiticity
of the bilocal operator
and the evenness of the polynomial from time reversal symmetry
of its matrix element.
Since we have observed complete Poincar\'{e} covariance
throughout the calculation, our model GPDs exhibit polynomiality.

Of particular interest are the cases $s=1$,
which reproduce partonic contributions (without charge weights) to
electromagnetic form factors (EMFFs),
and $s=2$, which give gravitational form factors (GFFs).
These relationships are given in Eqs.~(\ref{eqn:emffs},\ref{eqn:gffs}).

\begin{figure}
  \centering
  \includegraphics[width=\columnwidth]{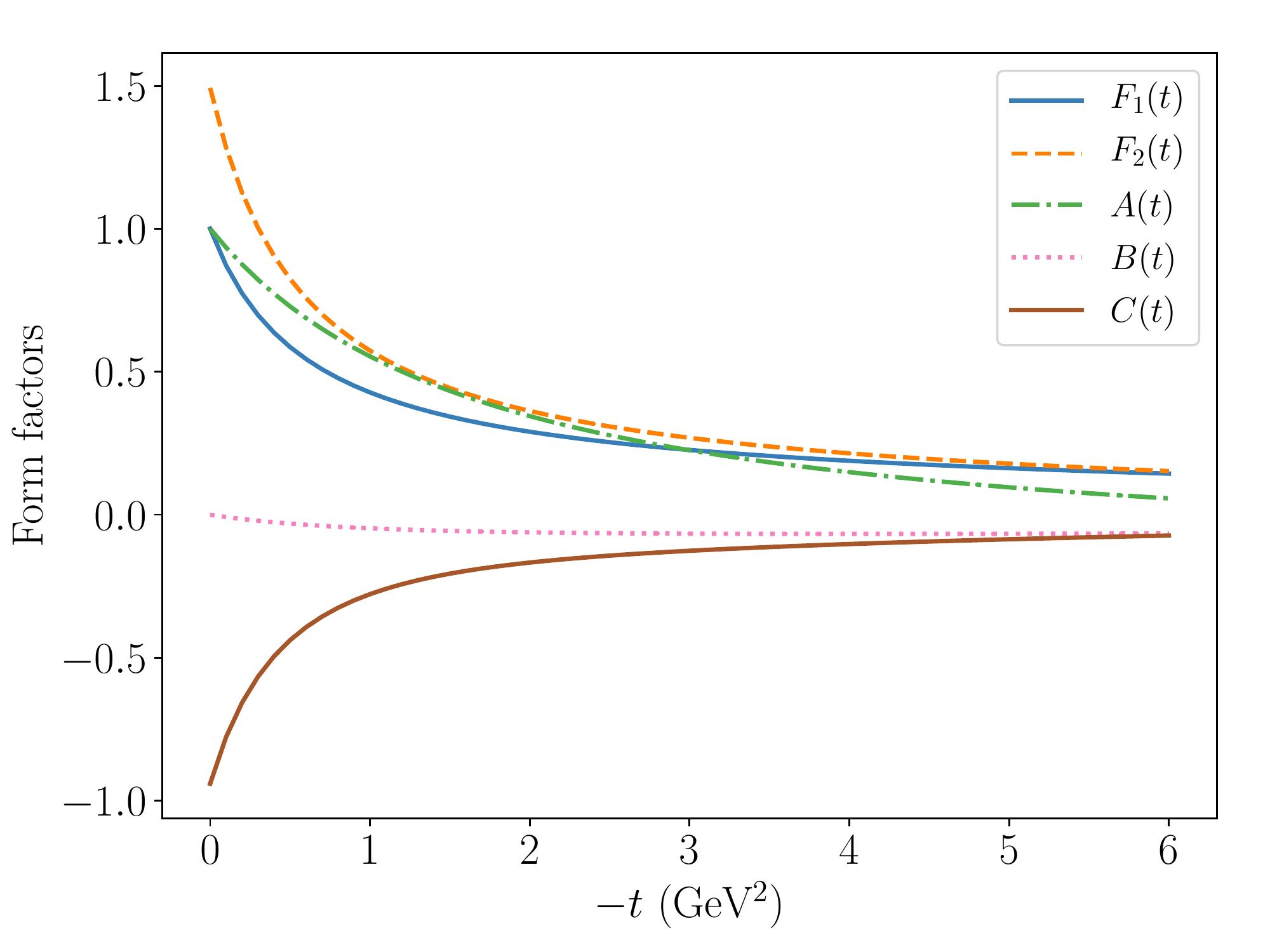}
  \caption{
    Electromagnetic and gravitational form factors of the proton
    as extracted from the leading-twist, helicity-independent GPDs.
  }
  \label{fig:ffs}
\end{figure}

The EMFFs of the proton have been previously calculated within the NJL model
(see, e.g., Ref.~\cite{Cloet:2014rja}),
but the GFFs have not been.
In Fig.~\ref{fig:ffs}, we present numerical results for the EMFFs and GFFs
extracted from the GPDs calculated in this work.
Since pion loops have not been included in this calculation,
our EMFFs should be compared to the ``bse'' results from Ref.~\cite{Cloet:2014rja}.

There are several constraints that the form factors must obey due to symmetries
and conservation laws.
Charge and momentum conservation require $F_1(0)=1$ and $A(0)=1$, respectively,
and our results satisfy these relations.
Conservation of angular momentum gives the Ji sum rule~\cite{Ji:1996ek}
$\frac{1}{2}\big(A(0)+B(0)\big) = \frac{1}{2}$,
or equivalently (when combined with momentum conservation) $B(0)=0$---a
statement otherwise known as the vanishing of the anomalous gravitomagnetic
moment.
We find that indeed $B(0)=0$ within our results,
and remark that this is an inevitable consequence of observing
Lorentz covariance throughout the calculation.

The remaining static observables $F_2(0)$ and $C(0)$ are not constrained
by symmetries or conservation laws.
The first of these gives the anomalous magnetic moment of the proton.
Empirically, $F_2(0) = \kappa_p = 1.793$,
but we underestimate this, finding $F_2(0) = 1.39$.
It was observed in Ref.~\cite{Cloet:2014rja} that perturbatively introducing
a pion cloud can significantly close the gap between the calculated
and model values.
The second of these, $C(0)$, is commonly known as the ``D-term''
\cite{Polyakov:1999gs},
and is neither constrained by symmetries nor by experiment.\footnote{
  There does however exist a recent model-dependent
  phenomenological extraction of the quark contribution
  from JLab Hall B data \cite{Burkert:2018bqq,Kumericki:2019ddg}.
}
We find $C(0)=-1.08$.

The form factor $C(t)$ has been interpreted as describing the distribution
of forces within hadrons \cite{Polyakov:2018zvc},
and the fact that $C(0) < 0$ is understood as an important stability criterion.
Since $C(0)$ is not constrained by any symmetries,
it is possible for operator dressing to change its value---in contrast to
$A(0)$ or $F_1(0)$.
In Ref.~\cite{Freese:2019bhb}, $C_\pi(t)$ was found to change by
a factor of $\sim3$ from dressing the quark-graviton vertex,
and introducing the dressing was necessary to satisfy a low-energy pion theorem.
In a similar vain, we find that dressing the nonlocal correlator
via the BSE depicted in Fig.~\ref{fig:bse} is necessary for proton stability.
If the bare nonlocal operator is used to calculate the proton GPDs,
then we obtain $C(0) = 0.85 > 0$, at stark odds with the apparent
stability of the proton.

To understand the necessity of operator dressing for observing proton stability,
we reiterate that dressing can be seen as accounting for the elementary current
quark substructure of the three dressed quarks that comprise the proton.
In the absence of dressing,
we would be attempting to describe the proton as being made of three
current quarks, which would not be accurate, and would account for only the
leading Fock state of the proton.
That we get $C(0) > 0$ in this case tells us that a hypothetical hadron
with the mass and quantum numbers of the proton that is made of only three
elementary current quarks would not be mechanically stable
in this framework.

Since the NJL model contains only quarks and $C(0)$ is not constrained
by any symmetries, it is possible that $C(0)$ may change significantly
with the introduction of gluons.
Therefore, our finding of $C(0)=-1.08$ should at best be interpreted as
a prediction for the quark contribution to the D-term,
rather than for the overall D-term of the proton.

~\\


\section{Summary and outlook}

We have calculated the helicity-independent, leading-twist GPDs of the proton
at finite skewness, in a confining version of the NJL model.
Dressing of non-local operator defining the light cone correlator---which
happens as a result of DCSB---was
necessary for sensible results to be obtained,
including a negative D-term compatible with the stability of the proton.
The Lorentz covariance of the model and all approximations made ensured
that polynomiality and sum rules relating to charge, energy-momentum,
and angular momentum conservation were all obeyed,
giving the vanishing of the anomalous
gravitomagnetic moment as a corollary.

It will be possible in future work to apply the same formalism to the
helicity-dependent and helicity-flip GPDs of the proton.
Moreover, the Lagrangian can be generalized to include immersion in
a finite-density medium, allowing predictions of GPDs in nuclear matter
and predictions for a generalized EMC effect.

\section*{Acknowledgments}

We would like to thank
Chao Shi
for illuminating discussions that helped contribute to our investigation.
This work was supported by the U.S.\ Department of Energy, Office of Science,
Office of Nuclear Physics, contract no.\ DE-AC02-06CH11357,
and an LDRD initiative at Argonne National Laboratory
under Project No.\ 2017-058-N0.


\bibliography{main.bib}

 \newcommand{\noop}[1]{}
\begin{thebibliography}{57}%
\makeatletter
\providecommand \@ifxundefined [1]{%
 \@ifx{#1\undefined}
}%
\providecommand \@ifnum [1]{%
 \ifnum #1\expandafter \@firstoftwo
 \else \expandafter \@secondoftwo
 \fi
}%
\providecommand \@ifx [1]{%
 \ifx #1\expandafter \@firstoftwo
 \else \expandafter \@secondoftwo
 \fi
}%
\providecommand \natexlab [1]{#1}%
\providecommand \enquote  [1]{``#1''}%
\providecommand \bibnamefont  [1]{#1}%
\providecommand \bibfnamefont [1]{#1}%
\providecommand \citenamefont [1]{#1}%
\providecommand \href@noop [0]{\@secondoftwo}%
\providecommand \href [0]{\begingroup \@sanitize@url \@href}%
\providecommand \@href[1]{\@@startlink{#1}\@@href}%
\providecommand \@@href[1]{\endgroup#1\@@endlink}%
\providecommand \@sanitize@url [0]{\catcode `\\12\catcode `\$12\catcode
  `\&12\catcode `\#12\catcode `\^12\catcode `\_12\catcode `\%12\relax}%
\providecommand \@@startlink[1]{}%
\providecommand \@@endlink[0]{}%
\providecommand \url  [0]{\begingroup\@sanitize@url \@url }%
\providecommand \@url [1]{\endgroup\@href {#1}{\urlprefix }}%
\providecommand \urlprefix  [0]{URL }%
\providecommand \Eprint [0]{\href }%
\providecommand \doibase [0]{http://dx.doi.org/}%
\providecommand \selectlanguage [0]{\@gobble}%
\providecommand \bibinfo  [0]{\@secondoftwo}%
\providecommand \bibfield  [0]{\@secondoftwo}%
\providecommand \translation [1]{[#1]}%
\providecommand \BibitemOpen [0]{}%
\providecommand \bibitemStop [0]{}%
\providecommand \bibitemNoStop [0]{.\EOS\space}%
\providecommand \EOS [0]{\spacefactor3000\relax}%
\providecommand \BibitemShut  [1]{\csname bibitem#1\endcsname}%
\let\auto@bib@innerbib\@empty
\bibitem [{\citenamefont {Collins}\ \emph {et~al.}(1997)\citenamefont
  {Collins}, \citenamefont {Frankfurt},\ and\ \citenamefont
  {Strikman}}]{Collins:1996fb}%
  \BibitemOpen
  \bibfield  {author} {\bibinfo {author} {\bibfnamefont {J.~C.}\ \bibnamefont
  {Collins}}, \bibinfo {author} {\bibfnamefont {L.}~\bibnamefont {Frankfurt}},
  \ and\ \bibinfo {author} {\bibfnamefont {M.}~\bibnamefont {Strikman}},\
  }\href {\doibase 10.1103/PhysRevD.56.2982} {\bibfield  {journal} {\bibinfo
  {journal} {Phys. Rev.}\ }\textbf {\bibinfo {volume} {D56}},\ \bibinfo {pages}
  {2982} (\bibinfo {year} {1997})},\ \Eprint
  {http://arxiv.org/abs/hep-ph/9611433} {arXiv:hep-ph/9611433 [hep-ph]}
  \BibitemShut {NoStop}%
\bibitem [{\citenamefont {Collins}\ and\ \citenamefont
  {Freund}(1999)}]{Collins:1998be}%
  \BibitemOpen
  \bibfield  {author} {\bibinfo {author} {\bibfnamefont {J.~C.}\ \bibnamefont
  {Collins}}\ and\ \bibinfo {author} {\bibfnamefont {A.}~\bibnamefont
  {Freund}},\ }\href {\doibase 10.1103/PhysRevD.59.074009} {\bibfield
  {journal} {\bibinfo  {journal} {Phys. Rev.}\ }\textbf {\bibinfo {volume}
  {D59}},\ \bibinfo {pages} {074009} (\bibinfo {year} {1999})},\ \Eprint
  {http://arxiv.org/abs/hep-ph/9801262} {arXiv:hep-ph/9801262 [hep-ph]}
  \BibitemShut {NoStop}%
\bibitem [{\citenamefont {Ji}\ and\ \citenamefont {Osborne}(1998)}]{Ji:1998xh}%
  \BibitemOpen
  \bibfield  {author} {\bibinfo {author} {\bibfnamefont {X.-D.}\ \bibnamefont
  {Ji}}\ and\ \bibinfo {author} {\bibfnamefont {J.}~\bibnamefont {Osborne}},\
  }\href {\doibase 10.1103/PhysRevD.58.094018} {\bibfield  {journal} {\bibinfo
  {journal} {Phys. Rev.}\ }\textbf {\bibinfo {volume} {D58}},\ \bibinfo {pages}
  {094018} (\bibinfo {year} {1998})},\ \Eprint
  {http://arxiv.org/abs/hep-ph/9801260} {arXiv:hep-ph/9801260 [hep-ph]}
  \BibitemShut {NoStop}%
\bibitem [{\citenamefont {Burkardt}(2003)}]{Burkardt:2002hr}%
  \BibitemOpen
  \bibfield  {author} {\bibinfo {author} {\bibfnamefont {M.}~\bibnamefont
  {Burkardt}},\ }\href {\doibase 10.1142/S0217751X03012370} {\bibfield
  {journal} {\bibinfo  {journal} {Int. J. Mod. Phys.}\ }\textbf {\bibinfo
  {volume} {A18}},\ \bibinfo {pages} {173} (\bibinfo {year} {2003})},\ \Eprint
  {http://arxiv.org/abs/hep-ph/0207047} {arXiv:hep-ph/0207047 [hep-ph]}
  \BibitemShut {NoStop}%
\bibitem [{\citenamefont {Lorcé}\ \emph {et~al.}(2019)\citenamefont {Lorcé},
  \citenamefont {Moutarde},\ and\ \citenamefont {Trawiński}}]{Lorce:2018egm}%
  \BibitemOpen
  \bibfield  {author} {\bibinfo {author} {\bibfnamefont {C.}~\bibnamefont
  {Lorcé}}, \bibinfo {author} {\bibfnamefont {H.}~\bibnamefont {Moutarde}}, \
  and\ \bibinfo {author} {\bibfnamefont {A.~P.}\ \bibnamefont {Trawiński}},\
  }\href {\doibase 10.1140/epjc/s10052-019-6572-3} {\bibfield  {journal}
  {\bibinfo  {journal} {Eur. Phys. J.}\ }\textbf {\bibinfo {volume} {C79}},\
  \bibinfo {pages} {89} (\bibinfo {year} {2019})},\ \Eprint
  {http://arxiv.org/abs/1810.09837} {arXiv:1810.09837 [hep-ph]} \BibitemShut
  {NoStop}%
\bibitem [{\citenamefont {Hatta}\ \emph {et~al.}(2018)\citenamefont {Hatta},
  \citenamefont {Rajan},\ and\ \citenamefont {Tanaka}}]{Hatta:2018sqd}%
  \BibitemOpen
  \bibfield  {author} {\bibinfo {author} {\bibfnamefont {Y.}~\bibnamefont
  {Hatta}}, \bibinfo {author} {\bibfnamefont {A.}~\bibnamefont {Rajan}}, \ and\
  \bibinfo {author} {\bibfnamefont {K.}~\bibnamefont {Tanaka}},\ }\href
  {\doibase 10.1007/JHEP12(2018)008} {\bibfield  {journal} {\bibinfo  {journal}
  {JHEP}\ }\textbf {\bibinfo {volume} {12}},\ \bibinfo {pages} {008} (\bibinfo
  {year} {2018})},\ \Eprint {http://arxiv.org/abs/1810.05116} {arXiv:1810.05116
  [hep-ph]} \BibitemShut {NoStop}%
\bibitem [{\citenamefont {Leader}\ and\ \citenamefont
  {Lorcé}(2014)}]{Leader:2013jra}%
  \BibitemOpen
  \bibfield  {author} {\bibinfo {author} {\bibfnamefont {E.}~\bibnamefont
  {Leader}}\ and\ \bibinfo {author} {\bibfnamefont {C.}~\bibnamefont
  {Lorcé}},\ }\href {\doibase 10.1016/j.physrep.2014.02.010} {\bibfield
  {journal} {\bibinfo  {journal} {Phys. Rept.}\ }\textbf {\bibinfo {volume}
  {541}},\ \bibinfo {pages} {163} (\bibinfo {year} {2014})},\ \Eprint
  {http://arxiv.org/abs/1309.4235} {arXiv:1309.4235 [hep-ph]} \BibitemShut
  {NoStop}%
\bibitem [{\citenamefont {Ji}(1997{\natexlab{a}})}]{Ji:1996ek}%
  \BibitemOpen
  \bibfield  {author} {\bibinfo {author} {\bibfnamefont {X.-D.}\ \bibnamefont
  {Ji}},\ }\href {\doibase 10.1103/PhysRevLett.78.610} {\bibfield  {journal}
  {\bibinfo  {journal} {Phys. Rev. Lett.}\ }\textbf {\bibinfo {volume} {78}},\
  \bibinfo {pages} {610} (\bibinfo {year} {1997}{\natexlab{a}})},\ \Eprint
  {http://arxiv.org/abs/hep-ph/9603249} {arXiv:hep-ph/9603249 [hep-ph]}
  \BibitemShut {NoStop}%
\bibitem [{\citenamefont {Polyakov}\ and\ \citenamefont
  {Schweitzer}(2018)}]{Polyakov:2018zvc}%
  \BibitemOpen
  \bibfield  {author} {\bibinfo {author} {\bibfnamefont {M.~V.}\ \bibnamefont
  {Polyakov}}\ and\ \bibinfo {author} {\bibfnamefont {P.}~\bibnamefont
  {Schweitzer}},\ }\href {\doibase 10.1142/S0217751X18300259} {\bibfield
  {journal} {\bibinfo  {journal} {Int. J. Mod. Phys.}\ }\textbf {\bibinfo
  {volume} {A33}},\ \bibinfo {pages} {1830025} (\bibinfo {year} {2018})},\
  \Eprint {http://arxiv.org/abs/1805.06596} {arXiv:1805.06596 [hep-ph]}
  \BibitemShut {NoStop}%
\bibitem [{\citenamefont {Freese}\ and\ \citenamefont
  {Cloët}(2019)}]{Freese:2019bhb}%
  \BibitemOpen
  \bibfield  {author} {\bibinfo {author} {\bibfnamefont {A.}~\bibnamefont
  {Freese}}\ and\ \bibinfo {author} {\bibfnamefont {I.~C.}\ \bibnamefont
  {Cloët}},\ }\href@noop {} {\  (\bibinfo {year} {2019})},\ \Eprint
  {http://arxiv.org/abs/1903.09222} {arXiv:1903.09222 [nucl-th]} \BibitemShut
  {NoStop}%
\bibitem [{\citenamefont {Vogl}\ and\ \citenamefont
  {Weise}(1991)}]{Vogl:1991qt}%
  \BibitemOpen
  \bibfield  {author} {\bibinfo {author} {\bibfnamefont {U.}~\bibnamefont
  {Vogl}}\ and\ \bibinfo {author} {\bibfnamefont {W.}~\bibnamefont {Weise}},\
  }\href {\doibase 10.1016/0146-6410(91)90005-9} {\bibfield  {journal}
  {\bibinfo  {journal} {Prog. Part. Nucl. Phys.}\ }\textbf {\bibinfo {volume}
  {27}},\ \bibinfo {pages} {195} (\bibinfo {year} {1991})}\BibitemShut
  {NoStop}%
\bibitem [{\citenamefont {Klevansky}(1992)}]{Klevansky:1992qe}%
  \BibitemOpen
  \bibfield  {author} {\bibinfo {author} {\bibfnamefont {S.~P.}\ \bibnamefont
  {Klevansky}},\ }\href {\doibase 10.1103/RevModPhys.64.649} {\bibfield
  {journal} {\bibinfo  {journal} {Rev. Mod. Phys.}\ }\textbf {\bibinfo {volume}
  {64}},\ \bibinfo {pages} {649} (\bibinfo {year} {1992})}\BibitemShut
  {NoStop}%
\bibitem [{\citenamefont {Hatsuda}\ and\ \citenamefont
  {Kunihiro}(1994)}]{Hatsuda:1994pi}%
  \BibitemOpen
  \bibfield  {author} {\bibinfo {author} {\bibfnamefont {T.}~\bibnamefont
  {Hatsuda}}\ and\ \bibinfo {author} {\bibfnamefont {T.}~\bibnamefont
  {Kunihiro}},\ }\href {\doibase 10.1016/0370-1573(94)90022-1} {\bibfield
  {journal} {\bibinfo  {journal} {Phys. Rept.}\ }\textbf {\bibinfo {volume}
  {247}},\ \bibinfo {pages} {221} (\bibinfo {year} {1994})},\ \Eprint
  {http://arxiv.org/abs/hep-ph/9401310} {arXiv:hep-ph/9401310 [hep-ph]}
  \BibitemShut {NoStop}%
\bibitem [{\citenamefont {Ebert}\ \emph {et~al.}(1996)\citenamefont {Ebert},
  \citenamefont {Feldmann},\ and\ \citenamefont {Reinhardt}}]{Ebert:1996vx}%
  \BibitemOpen
  \bibfield  {author} {\bibinfo {author} {\bibfnamefont {D.}~\bibnamefont
  {Ebert}}, \bibinfo {author} {\bibfnamefont {T.}~\bibnamefont {Feldmann}}, \
  and\ \bibinfo {author} {\bibfnamefont {H.}~\bibnamefont {Reinhardt}},\ }\href
  {\doibase 10.1016/0370-2693(96)01158-6} {\bibfield  {journal} {\bibinfo
  {journal} {Phys. Lett.}\ }\textbf {\bibinfo {volume} {B388}},\ \bibinfo
  {pages} {154} (\bibinfo {year} {1996})},\ \Eprint
  {http://arxiv.org/abs/hep-ph/9608223} {arXiv:hep-ph/9608223 [hep-ph]}
  \BibitemShut {NoStop}%
\bibitem [{\citenamefont {Hellstern}\ \emph {et~al.}(1997)\citenamefont
  {Hellstern}, \citenamefont {Alkofer},\ and\ \citenamefont
  {Reinhardt}}]{Hellstern:1997nv}%
  \BibitemOpen
  \bibfield  {author} {\bibinfo {author} {\bibfnamefont {G.}~\bibnamefont
  {Hellstern}}, \bibinfo {author} {\bibfnamefont {R.}~\bibnamefont {Alkofer}},
  \ and\ \bibinfo {author} {\bibfnamefont {H.}~\bibnamefont {Reinhardt}},\
  }\href {\doibase 10.1016/S0375-9474(97)00412-0} {\bibfield  {journal}
  {\bibinfo  {journal} {Nucl. Phys.}\ }\textbf {\bibinfo {volume} {A625}},\
  \bibinfo {pages} {697} (\bibinfo {year} {1997})},\ \Eprint
  {http://arxiv.org/abs/hep-ph/9706551} {arXiv:hep-ph/9706551 [hep-ph]}
  \BibitemShut {NoStop}%
\bibitem [{\citenamefont {Cloët}\ \emph {et~al.}(2014)\citenamefont {Cloët},
  \citenamefont {Bentz},\ and\ \citenamefont {Thomas}}]{Cloet:2014rja}%
  \BibitemOpen
  \bibfield  {author} {\bibinfo {author} {\bibfnamefont {I.~C.}\ \bibnamefont
  {Cloët}}, \bibinfo {author} {\bibfnamefont {W.}~\bibnamefont {Bentz}}, \
  and\ \bibinfo {author} {\bibfnamefont {A.~W.}\ \bibnamefont {Thomas}},\
  }\href {\doibase 10.1103/PhysRevC.90.045202} {\bibfield  {journal} {\bibinfo
  {journal} {Phys. Rev.}\ }\textbf {\bibinfo {volume} {C90}},\ \bibinfo {pages}
  {045202} (\bibinfo {year} {2014})},\ \Eprint {http://arxiv.org/abs/1405.5542}
  {arXiv:1405.5542 [nucl-th]} \BibitemShut {NoStop}%
\bibitem [{\citenamefont {Ninomiya}\ \emph {et~al.}(2017)\citenamefont
  {Ninomiya}, \citenamefont {Bentz},\ and\ \citenamefont
  {Cloët}}]{Ninomiya:2017ggn}%
  \BibitemOpen
  \bibfield  {author} {\bibinfo {author} {\bibfnamefont {Y.}~\bibnamefont
  {Ninomiya}}, \bibinfo {author} {\bibfnamefont {W.}~\bibnamefont {Bentz}}, \
  and\ \bibinfo {author} {\bibfnamefont {I.~C.}\ \bibnamefont {Cloët}},\
  }\href {\doibase 10.1103/PhysRevC.96.045206} {\bibfield  {journal} {\bibinfo
  {journal} {Phys. Rev.}\ }\textbf {\bibinfo {volume} {C96}},\ \bibinfo {pages}
  {045206} (\bibinfo {year} {2017})},\ \Eprint
  {http://arxiv.org/abs/1707.03787} {arXiv:1707.03787 [nucl-th]} \BibitemShut
  {NoStop}%
\bibitem [{\citenamefont {Ishii}\ \emph
  {et~al.}(1993{\natexlab{a}})\citenamefont {Ishii}, \citenamefont {Bentz},\
  and\ \citenamefont {Yazaki}}]{Ishii:1993np}%
  \BibitemOpen
  \bibfield  {author} {\bibinfo {author} {\bibfnamefont {N.}~\bibnamefont
  {Ishii}}, \bibinfo {author} {\bibfnamefont {W.}~\bibnamefont {Bentz}}, \ and\
  \bibinfo {author} {\bibfnamefont {K.}~\bibnamefont {Yazaki}},\ }\href
  {\doibase 10.1016/0370-2693(93)90683-9} {\bibfield  {journal} {\bibinfo
  {journal} {Phys. Lett.}\ }\textbf {\bibinfo {volume} {B301}},\ \bibinfo
  {pages} {165} (\bibinfo {year} {1993}{\natexlab{a}})}\BibitemShut {NoStop}%
\bibitem [{\citenamefont {Ishii}\ \emph
  {et~al.}(1993{\natexlab{b}})\citenamefont {Ishii}, \citenamefont {Bentz},\
  and\ \citenamefont {Yazaki}}]{Ishii:1993rt}%
  \BibitemOpen
  \bibfield  {author} {\bibinfo {author} {\bibfnamefont {N.}~\bibnamefont
  {Ishii}}, \bibinfo {author} {\bibfnamefont {W.}~\bibnamefont {Bentz}}, \ and\
  \bibinfo {author} {\bibfnamefont {K.}~\bibnamefont {Yazaki}},\ }\href
  {\doibase 10.1016/0370-2693(93)91778-L} {\bibfield  {journal} {\bibinfo
  {journal} {Phys. Lett.}\ }\textbf {\bibinfo {volume} {B318}},\ \bibinfo
  {pages} {26} (\bibinfo {year} {1993}{\natexlab{b}})}\BibitemShut {NoStop}%
\bibitem [{\citenamefont {Ishii}\ \emph {et~al.}(1995)\citenamefont {Ishii},
  \citenamefont {Bentz},\ and\ \citenamefont {Yazaki}}]{Ishii:1995bu}%
  \BibitemOpen
  \bibfield  {author} {\bibinfo {author} {\bibfnamefont {N.}~\bibnamefont
  {Ishii}}, \bibinfo {author} {\bibfnamefont {W.}~\bibnamefont {Bentz}}, \ and\
  \bibinfo {author} {\bibfnamefont {K.}~\bibnamefont {Yazaki}},\ }\href
  {\doibase 10.1016/0375-9474(95)00032-V} {\bibfield  {journal} {\bibinfo
  {journal} {Nucl. Phys.}\ }\textbf {\bibinfo {volume} {A587}},\ \bibinfo
  {pages} {617} (\bibinfo {year} {1995})}\BibitemShut {NoStop}%
\bibitem [{\citenamefont {Cahill}\ \emph {et~al.}(1989)\citenamefont {Cahill},
  \citenamefont {Roberts},\ and\ \citenamefont {Praschifka}}]{Cahill:1988dx}%
  \BibitemOpen
  \bibfield  {author} {\bibinfo {author} {\bibfnamefont {R.~T.}\ \bibnamefont
  {Cahill}}, \bibinfo {author} {\bibfnamefont {C.~D.}\ \bibnamefont {Roberts}},
  \ and\ \bibinfo {author} {\bibfnamefont {J.}~\bibnamefont {Praschifka}},\
  }\href {\doibase 10.1071/PH890129} {\bibfield  {journal} {\bibinfo  {journal}
  {Austral. J. Phys.}\ }\textbf {\bibinfo {volume} {42}},\ \bibinfo {pages}
  {129} (\bibinfo {year} {1989})}\BibitemShut {NoStop}%
\bibitem [{\citenamefont {Anselmino}\ \emph {et~al.}(1993)\citenamefont
  {Anselmino}, \citenamefont {Predazzi}, \citenamefont {Ekelin}, \citenamefont
  {Fredriksson},\ and\ \citenamefont {Lichtenberg}}]{Anselmino:1992vg}%
  \BibitemOpen
  \bibfield  {author} {\bibinfo {author} {\bibfnamefont {M.}~\bibnamefont
  {Anselmino}}, \bibinfo {author} {\bibfnamefont {E.}~\bibnamefont {Predazzi}},
  \bibinfo {author} {\bibfnamefont {S.}~\bibnamefont {Ekelin}}, \bibinfo
  {author} {\bibfnamefont {S.}~\bibnamefont {Fredriksson}}, \ and\ \bibinfo
  {author} {\bibfnamefont {D.~B.}\ \bibnamefont {Lichtenberg}},\ }\href
  {\doibase 10.1103/RevModPhys.65.1199} {\bibfield  {journal} {\bibinfo
  {journal} {Rev. Mod. Phys.}\ }\textbf {\bibinfo {volume} {65}},\ \bibinfo
  {pages} {1199} (\bibinfo {year} {1993})}\BibitemShut {NoStop}%
\bibitem [{\citenamefont {Roberts}\ \emph {et~al.}(2011)\citenamefont
  {Roberts}, \citenamefont {Chang}, \citenamefont {Cloet},\ and\ \citenamefont
  {Roberts}}]{Roberts:2011cf}%
  \BibitemOpen
  \bibfield  {author} {\bibinfo {author} {\bibfnamefont {H.~L.~L.}\
  \bibnamefont {Roberts}}, \bibinfo {author} {\bibfnamefont {L.}~\bibnamefont
  {Chang}}, \bibinfo {author} {\bibfnamefont {I.~C.}\ \bibnamefont {Cloet}}, \
  and\ \bibinfo {author} {\bibfnamefont {C.~D.}\ \bibnamefont {Roberts}},\
  }\href {\doibase 10.1007/s00601-011-0225-x} {\bibfield  {journal} {\bibinfo
  {journal} {Few Body Syst.}\ }\textbf {\bibinfo {volume} {51}},\ \bibinfo
  {pages} {1} (\bibinfo {year} {2011})},\ \Eprint
  {http://arxiv.org/abs/1101.4244} {arXiv:1101.4244 [nucl-th]} \BibitemShut
  {NoStop}%
\bibitem [{\citenamefont {Cates}\ \emph {et~al.}(2011)\citenamefont {Cates},
  \citenamefont {de~Jager}, \citenamefont {Riordan},\ and\ \citenamefont
  {Wojtsekhowski}}]{Cates:2011pz}%
  \BibitemOpen
  \bibfield  {author} {\bibinfo {author} {\bibfnamefont {G.~D.}\ \bibnamefont
  {Cates}}, \bibinfo {author} {\bibfnamefont {C.~W.}\ \bibnamefont {de~Jager}},
  \bibinfo {author} {\bibfnamefont {S.}~\bibnamefont {Riordan}}, \ and\
  \bibinfo {author} {\bibfnamefont {B.}~\bibnamefont {Wojtsekhowski}},\ }\href
  {\doibase 10.1103/PhysRevLett.106.252003} {\bibfield  {journal} {\bibinfo
  {journal} {Phys. Rev. Lett.}\ }\textbf {\bibinfo {volume} {106}},\ \bibinfo
  {pages} {252003} (\bibinfo {year} {2011})},\ \Eprint
  {http://arxiv.org/abs/1103.1808} {arXiv:1103.1808 [nucl-ex]} \BibitemShut
  {NoStop}%
\bibitem [{\citenamefont {Brodsky}\ \emph {et~al.}(2016)\citenamefont
  {Brodsky}, \citenamefont {de~Téramond}, \citenamefont {Dosch},\ and\
  \citenamefont {Lorcé}}]{Brodsky:2016rvj}%
  \BibitemOpen
  \bibfield  {author} {\bibinfo {author} {\bibfnamefont {S.~J.}\ \bibnamefont
  {Brodsky}}, \bibinfo {author} {\bibfnamefont {G.~F.}\ \bibnamefont
  {de~Téramond}}, \bibinfo {author} {\bibfnamefont {H.~G.}\ \bibnamefont
  {Dosch}}, \ and\ \bibinfo {author} {\bibfnamefont {C.}~\bibnamefont
  {Lorcé}},\ }\href {\doibase 10.1142/S0217751X16300295} {\bibfield  {journal}
  {\bibinfo  {journal} {Int. J. Mod. Phys.}\ }\textbf {\bibinfo {volume}
  {A31}},\ \bibinfo {pages} {1630029} (\bibinfo {year} {2016})},\ \Eprint
  {http://arxiv.org/abs/1606.04638} {arXiv:1606.04638 [hep-ph]} \BibitemShut
  {NoStop}%
\bibitem [{\citenamefont {Mineo}\ \emph {et~al.}(1999)\citenamefont {Mineo},
  \citenamefont {Bentz},\ and\ \citenamefont {Yazaki}}]{Mineo:1999eq}%
  \BibitemOpen
  \bibfield  {author} {\bibinfo {author} {\bibfnamefont {H.}~\bibnamefont
  {Mineo}}, \bibinfo {author} {\bibfnamefont {W.}~\bibnamefont {Bentz}}, \ and\
  \bibinfo {author} {\bibfnamefont {K.}~\bibnamefont {Yazaki}},\ }\href
  {\doibase 10.1103/PhysRevC.60.065201} {\bibfield  {journal} {\bibinfo
  {journal} {Phys. Rev.}\ }\textbf {\bibinfo {volume} {C60}},\ \bibinfo {pages}
  {065201} (\bibinfo {year} {1999})},\ \Eprint
  {http://arxiv.org/abs/nucl-th/9907043} {arXiv:nucl-th/9907043 [nucl-th]}
  \BibitemShut {NoStop}%
\bibitem [{\citenamefont {Cloet}\ \emph
  {et~al.}(2005{\natexlab{a}})\citenamefont {Cloet}, \citenamefont {Bentz},\
  and\ \citenamefont {Thomas}}]{Cloet:2005rt}%
  \BibitemOpen
  \bibfield  {author} {\bibinfo {author} {\bibfnamefont {I.~C.}\ \bibnamefont
  {Cloet}}, \bibinfo {author} {\bibfnamefont {W.}~\bibnamefont {Bentz}}, \ and\
  \bibinfo {author} {\bibfnamefont {A.~W.}\ \bibnamefont {Thomas}},\ }\href
  {\doibase 10.1103/PhysRevLett.95.052302} {\bibfield  {journal} {\bibinfo
  {journal} {Phys. Rev. Lett.}\ }\textbf {\bibinfo {volume} {95}},\ \bibinfo
  {pages} {052302} (\bibinfo {year} {2005}{\natexlab{a}})},\ \Eprint
  {http://arxiv.org/abs/nucl-th/0504019} {arXiv:nucl-th/0504019 [nucl-th]}
  \BibitemShut {NoStop}%
\bibitem [{\citenamefont {Cloet}\ \emph
  {et~al.}(2005{\natexlab{b}})\citenamefont {Cloet}, \citenamefont {Bentz},\
  and\ \citenamefont {Thomas}}]{Cloet:2005pp}%
  \BibitemOpen
  \bibfield  {author} {\bibinfo {author} {\bibfnamefont {I.~C.}\ \bibnamefont
  {Cloet}}, \bibinfo {author} {\bibfnamefont {W.}~\bibnamefont {Bentz}}, \ and\
  \bibinfo {author} {\bibfnamefont {A.~W.}\ \bibnamefont {Thomas}},\ }\href
  {\doibase 10.1016/j.physletb.2005.06.065} {\bibfield  {journal} {\bibinfo
  {journal} {Phys. Lett.}\ }\textbf {\bibinfo {volume} {B621}},\ \bibinfo
  {pages} {246} (\bibinfo {year} {2005}{\natexlab{b}})},\ \Eprint
  {http://arxiv.org/abs/hep-ph/0504229} {arXiv:hep-ph/0504229 [hep-ph]}
  \BibitemShut {NoStop}%
\bibitem [{\citenamefont {Mineo}\ \emph {et~al.}(2005)\citenamefont {Mineo},
  \citenamefont {Yang}, \citenamefont {Cheung},\ and\ \citenamefont
  {Bentz}}]{Mineo:2005vs}%
  \BibitemOpen
  \bibfield  {author} {\bibinfo {author} {\bibfnamefont {H.}~\bibnamefont
  {Mineo}}, \bibinfo {author} {\bibfnamefont {S.~N.}\ \bibnamefont {Yang}},
  \bibinfo {author} {\bibfnamefont {C.-Y.}\ \bibnamefont {Cheung}}, \ and\
  \bibinfo {author} {\bibfnamefont {W.}~\bibnamefont {Bentz}},\ }\href
  {\doibase 10.1103/PhysRevC.72.025202} {\bibfield  {journal} {\bibinfo
  {journal} {Phys. Rev.}\ }\textbf {\bibinfo {volume} {C72}},\ \bibinfo {pages}
  {025202} (\bibinfo {year} {2005})}\BibitemShut {NoStop}%
\bibitem [{\citenamefont {Cloet}\ \emph {et~al.}(2006)\citenamefont {Cloet},
  \citenamefont {Bentz},\ and\ \citenamefont {Thomas}}]{Cloet:2006bq}%
  \BibitemOpen
  \bibfield  {author} {\bibinfo {author} {\bibfnamefont {I.~C.}\ \bibnamefont
  {Cloet}}, \bibinfo {author} {\bibfnamefont {W.}~\bibnamefont {Bentz}}, \ and\
  \bibinfo {author} {\bibfnamefont {A.~W.}\ \bibnamefont {Thomas}},\ }\href
  {\doibase 10.1016/j.physletb.2006.08.076} {\bibfield  {journal} {\bibinfo
  {journal} {Phys. Lett.}\ }\textbf {\bibinfo {volume} {B642}},\ \bibinfo
  {pages} {210} (\bibinfo {year} {2006})},\ \Eprint
  {http://arxiv.org/abs/nucl-th/0605061} {arXiv:nucl-th/0605061 [nucl-th]}
  \BibitemShut {NoStop}%
\bibitem [{\citenamefont {Cloet}\ \emph
  {et~al.}(2008{\natexlab{a}})\citenamefont {Cloet}, \citenamefont {Bentz},\
  and\ \citenamefont {Thomas}}]{Cloet:2007em}%
  \BibitemOpen
  \bibfield  {author} {\bibinfo {author} {\bibfnamefont {I.~C.}\ \bibnamefont
  {Cloet}}, \bibinfo {author} {\bibfnamefont {W.}~\bibnamefont {Bentz}}, \ and\
  \bibinfo {author} {\bibfnamefont {A.~W.}\ \bibnamefont {Thomas}},\ }\href
  {\doibase 10.1016/j.physletb.2007.09.071} {\bibfield  {journal} {\bibinfo
  {journal} {Phys. Lett.}\ }\textbf {\bibinfo {volume} {B659}},\ \bibinfo
  {pages} {214} (\bibinfo {year} {2008}{\natexlab{a}})},\ \Eprint
  {http://arxiv.org/abs/0708.3246} {arXiv:0708.3246 [hep-ph]} \BibitemShut
  {NoStop}%
\bibitem [{\citenamefont {Eichmann}\ \emph {et~al.}(2008)\citenamefont
  {Eichmann}, \citenamefont {Krassnigg}, \citenamefont {Schwinzerl},\ and\
  \citenamefont {Alkofer}}]{Eichmann:2007nn}%
  \BibitemOpen
  \bibfield  {author} {\bibinfo {author} {\bibfnamefont {G.}~\bibnamefont
  {Eichmann}}, \bibinfo {author} {\bibfnamefont {A.}~\bibnamefont {Krassnigg}},
  \bibinfo {author} {\bibfnamefont {M.}~\bibnamefont {Schwinzerl}}, \ and\
  \bibinfo {author} {\bibfnamefont {R.}~\bibnamefont {Alkofer}},\ }\href
  {\doibase 10.1016/j.aop.2008.02.007} {\bibfield  {journal} {\bibinfo
  {journal} {Annals Phys.}\ }\textbf {\bibinfo {volume} {323}},\ \bibinfo
  {pages} {2505} (\bibinfo {year} {2008})},\ \Eprint
  {http://arxiv.org/abs/0712.2666} {arXiv:0712.2666 [hep-ph]} \BibitemShut
  {NoStop}%
\bibitem [{\citenamefont {Cloet}\ \emph
  {et~al.}(2008{\natexlab{b}})\citenamefont {Cloet}, \citenamefont {Eichmann},
  \citenamefont {Flambaum}, \citenamefont {Roberts}, \citenamefont {Bhagwat},\
  and\ \citenamefont {Holl}}]{Cloet:2008wg}%
  \BibitemOpen
  \bibfield  {author} {\bibinfo {author} {\bibfnamefont {I.~C.}\ \bibnamefont
  {Cloet}}, \bibinfo {author} {\bibfnamefont {G.}~\bibnamefont {Eichmann}},
  \bibinfo {author} {\bibfnamefont {V.~V.}\ \bibnamefont {Flambaum}}, \bibinfo
  {author} {\bibfnamefont {C.~D.}\ \bibnamefont {Roberts}}, \bibinfo {author}
  {\bibfnamefont {M.~S.}\ \bibnamefont {Bhagwat}}, \ and\ \bibinfo {author}
  {\bibfnamefont {A.}~\bibnamefont {Holl}},\ }\href {\doibase
  10.1007/s00601-008-0240-8} {\bibfield  {journal} {\bibinfo  {journal} {Few
  Body Syst.}\ }\textbf {\bibinfo {volume} {42}},\ \bibinfo {pages} {91}
  (\bibinfo {year} {2008}{\natexlab{b}})},\ \Eprint
  {http://arxiv.org/abs/0804.3118} {arXiv:0804.3118 [nucl-th]} \BibitemShut
  {NoStop}%
\bibitem [{\citenamefont {Eichmann}\ \emph {et~al.}(2009)\citenamefont
  {Eichmann}, \citenamefont {Cloet}, \citenamefont {Alkofer}, \citenamefont
  {Krassnigg},\ and\ \citenamefont {Roberts}}]{Eichmann:2008ef}%
  \BibitemOpen
  \bibfield  {author} {\bibinfo {author} {\bibfnamefont {G.}~\bibnamefont
  {Eichmann}}, \bibinfo {author} {\bibfnamefont {I.~C.}\ \bibnamefont {Cloet}},
  \bibinfo {author} {\bibfnamefont {R.}~\bibnamefont {Alkofer}}, \bibinfo
  {author} {\bibfnamefont {A.}~\bibnamefont {Krassnigg}}, \ and\ \bibinfo
  {author} {\bibfnamefont {C.~D.}\ \bibnamefont {Roberts}},\ }\href {\doibase
  10.1103/PhysRevC.79.012202} {\bibfield  {journal} {\bibinfo  {journal} {Phys.
  Rev.}\ }\textbf {\bibinfo {volume} {C79}},\ \bibinfo {pages} {012202}
  (\bibinfo {year} {2009})},\ \Eprint {http://arxiv.org/abs/0810.1222}
  {arXiv:0810.1222 [nucl-th]} \BibitemShut {NoStop}%
\bibitem [{\citenamefont {Nicmorus}\ \emph {et~al.}(2009)\citenamefont
  {Nicmorus}, \citenamefont {Eichmann}, \citenamefont {Krassnigg},\ and\
  \citenamefont {Alkofer}}]{Nicmorus:2008vb}%
  \BibitemOpen
  \bibfield  {author} {\bibinfo {author} {\bibfnamefont {D.}~\bibnamefont
  {Nicmorus}}, \bibinfo {author} {\bibfnamefont {G.}~\bibnamefont {Eichmann}},
  \bibinfo {author} {\bibfnamefont {A.}~\bibnamefont {Krassnigg}}, \ and\
  \bibinfo {author} {\bibfnamefont {R.}~\bibnamefont {Alkofer}},\ }\href
  {\doibase 10.1103/PhysRevD.80.054028} {\bibfield  {journal} {\bibinfo
  {journal} {Phys. Rev.}\ }\textbf {\bibinfo {volume} {D80}},\ \bibinfo {pages}
  {054028} (\bibinfo {year} {2009})},\ \Eprint {http://arxiv.org/abs/0812.1665}
  {arXiv:0812.1665 [hep-ph]} \BibitemShut {NoStop}%
\bibitem [{\citenamefont {Nicmorus}\ \emph {et~al.}(2010)\citenamefont
  {Nicmorus}, \citenamefont {Eichmann},\ and\ \citenamefont
  {Alkofer}}]{Nicmorus:2010sd}%
  \BibitemOpen
  \bibfield  {author} {\bibinfo {author} {\bibfnamefont {D.}~\bibnamefont
  {Nicmorus}}, \bibinfo {author} {\bibfnamefont {G.}~\bibnamefont {Eichmann}},
  \ and\ \bibinfo {author} {\bibfnamefont {R.}~\bibnamefont {Alkofer}},\ }\href
  {\doibase 10.1103/PhysRevD.82.114017} {\bibfield  {journal} {\bibinfo
  {journal} {Phys. Rev.}\ }\textbf {\bibinfo {volume} {D82}},\ \bibinfo {pages}
  {114017} (\bibinfo {year} {2010})},\ \Eprint {http://arxiv.org/abs/1008.3184}
  {arXiv:1008.3184 [hep-ph]} \BibitemShut {NoStop}%
\bibitem [{\citenamefont {Matevosyan}\ \emph {et~al.}(2012)\citenamefont
  {Matevosyan}, \citenamefont {Bentz}, \citenamefont {Cloet},\ and\
  \citenamefont {Thomas}}]{Matevosyan:2011vj}%
  \BibitemOpen
  \bibfield  {author} {\bibinfo {author} {\bibfnamefont {H.~H.}\ \bibnamefont
  {Matevosyan}}, \bibinfo {author} {\bibfnamefont {W.}~\bibnamefont {Bentz}},
  \bibinfo {author} {\bibfnamefont {I.~C.}\ \bibnamefont {Cloet}}, \ and\
  \bibinfo {author} {\bibfnamefont {A.~W.}\ \bibnamefont {Thomas}},\ }\href
  {\doibase 10.1103/PhysRevD.85.014021} {\bibfield  {journal} {\bibinfo
  {journal} {Phys. Rev.}\ }\textbf {\bibinfo {volume} {D85}},\ \bibinfo {pages}
  {014021} (\bibinfo {year} {2012})},\ \Eprint {http://arxiv.org/abs/1111.1740}
  {arXiv:1111.1740 [hep-ph]} \BibitemShut {NoStop}%
\bibitem [{\citenamefont {Wilson}\ \emph {et~al.}(2012)\citenamefont {Wilson},
  \citenamefont {Cloet}, \citenamefont {Chang},\ and\ \citenamefont
  {Roberts}}]{Wilson:2011aa}%
  \BibitemOpen
  \bibfield  {author} {\bibinfo {author} {\bibfnamefont {D.~J.}\ \bibnamefont
  {Wilson}}, \bibinfo {author} {\bibfnamefont {I.~C.}\ \bibnamefont {Cloet}},
  \bibinfo {author} {\bibfnamefont {L.}~\bibnamefont {Chang}}, \ and\ \bibinfo
  {author} {\bibfnamefont {C.~D.}\ \bibnamefont {Roberts}},\ }\href {\doibase
  10.1103/PhysRevC.85.025205} {\bibfield  {journal} {\bibinfo  {journal} {Phys.
  Rev.}\ }\textbf {\bibinfo {volume} {C85}},\ \bibinfo {pages} {025205}
  (\bibinfo {year} {2012})},\ \Eprint {http://arxiv.org/abs/1112.2212}
  {arXiv:1112.2212 [nucl-th]} \BibitemShut {NoStop}%
\bibitem [{\citenamefont {Cloet}\ and\ \citenamefont
  {Miller}(2012)}]{Cloet:2012cy}%
  \BibitemOpen
  \bibfield  {author} {\bibinfo {author} {\bibfnamefont {I.~C.}\ \bibnamefont
  {Cloet}}\ and\ \bibinfo {author} {\bibfnamefont {G.~A.}\ \bibnamefont
  {Miller}},\ }\href {\doibase 10.1103/PhysRevC.86.015208} {\bibfield
  {journal} {\bibinfo  {journal} {Phys. Rev.}\ }\textbf {\bibinfo {volume}
  {C86}},\ \bibinfo {pages} {015208} (\bibinfo {year} {2012})},\ \Eprint
  {http://arxiv.org/abs/1204.4422} {arXiv:1204.4422 [nucl-th]} \BibitemShut
  {NoStop}%
\bibitem [{\citenamefont {Cloet}\ \emph {et~al.}(2013)\citenamefont {Cloet},
  \citenamefont {Roberts},\ and\ \citenamefont {Thomas}}]{Cloet:2013gva}%
  \BibitemOpen
  \bibfield  {author} {\bibinfo {author} {\bibfnamefont {I.~C.}\ \bibnamefont
  {Cloet}}, \bibinfo {author} {\bibfnamefont {C.~D.}\ \bibnamefont {Roberts}},
  \ and\ \bibinfo {author} {\bibfnamefont {A.~W.}\ \bibnamefont {Thomas}},\
  }\href {\doibase 10.1103/PhysRevLett.111.101803} {\bibfield  {journal}
  {\bibinfo  {journal} {Phys. Rev. Lett.}\ }\textbf {\bibinfo {volume} {111}},\
  \bibinfo {pages} {101803} (\bibinfo {year} {2013})},\ \Eprint
  {http://arxiv.org/abs/1304.0855} {arXiv:1304.0855 [nucl-th]} \BibitemShut
  {NoStop}%
\bibitem [{\citenamefont {Segovia}\ \emph
  {et~al.}(2014{\natexlab{a}})\citenamefont {Segovia}, \citenamefont {Chen},
  \citenamefont {Cloët}, \citenamefont {Roberts}, \citenamefont {Schmidt},\
  and\ \citenamefont {Wan}}]{Segovia:2013uga}%
  \BibitemOpen
  \bibfield  {author} {\bibinfo {author} {\bibfnamefont {J.}~\bibnamefont
  {Segovia}}, \bibinfo {author} {\bibfnamefont {C.}~\bibnamefont {Chen}},
  \bibinfo {author} {\bibfnamefont {I.~C.}\ \bibnamefont {Cloët}}, \bibinfo
  {author} {\bibfnamefont {C.~D.}\ \bibnamefont {Roberts}}, \bibinfo {author}
  {\bibfnamefont {S.~M.}\ \bibnamefont {Schmidt}}, \ and\ \bibinfo {author}
  {\bibfnamefont {S.}~\bibnamefont {Wan}},\ }\href {\doibase
  10.1007/s00601-013-0734-x} {\bibfield  {journal} {\bibinfo  {journal} {Few
  Body Syst.}\ }\textbf {\bibinfo {volume} {55}},\ \bibinfo {pages} {1}
  (\bibinfo {year} {2014}{\natexlab{a}})},\ \Eprint
  {http://arxiv.org/abs/1308.5225} {arXiv:1308.5225 [nucl-th]} \BibitemShut
  {NoStop}%
\bibitem [{\citenamefont {Segovia}\ \emph
  {et~al.}(2014{\natexlab{b}})\citenamefont {Segovia}, \citenamefont {Cloet},
  \citenamefont {Roberts},\ and\ \citenamefont {Schmidt}}]{Segovia:2014aza}%
  \BibitemOpen
  \bibfield  {author} {\bibinfo {author} {\bibfnamefont {J.}~\bibnamefont
  {Segovia}}, \bibinfo {author} {\bibfnamefont {I.~C.}\ \bibnamefont {Cloet}},
  \bibinfo {author} {\bibfnamefont {C.~D.}\ \bibnamefont {Roberts}}, \ and\
  \bibinfo {author} {\bibfnamefont {S.~M.}\ \bibnamefont {Schmidt}},\ }\href
  {\doibase 10.1007/s00601-014-0907-2} {\bibfield  {journal} {\bibinfo
  {journal} {Few Body Syst.}\ }\textbf {\bibinfo {volume} {55}},\ \bibinfo
  {pages} {1185} (\bibinfo {year} {2014}{\natexlab{b}})},\ \Eprint
  {http://arxiv.org/abs/1408.2919} {arXiv:1408.2919 [nucl-th]} \BibitemShut
  {NoStop}%
\bibitem [{\citenamefont {Segovia}\ \emph {et~al.}(2015)\citenamefont
  {Segovia}, \citenamefont {El-Bennich}, \citenamefont {Rojas}, \citenamefont
  {Cloet}, \citenamefont {Roberts}, \citenamefont {Xu},\ and\ \citenamefont
  {Zong}}]{Segovia:2015hra}%
  \BibitemOpen
  \bibfield  {author} {\bibinfo {author} {\bibfnamefont {J.}~\bibnamefont
  {Segovia}}, \bibinfo {author} {\bibfnamefont {B.}~\bibnamefont {El-Bennich}},
  \bibinfo {author} {\bibfnamefont {E.}~\bibnamefont {Rojas}}, \bibinfo
  {author} {\bibfnamefont {I.~C.}\ \bibnamefont {Cloet}}, \bibinfo {author}
  {\bibfnamefont {C.~D.}\ \bibnamefont {Roberts}}, \bibinfo {author}
  {\bibfnamefont {S.-S.}\ \bibnamefont {Xu}}, \ and\ \bibinfo {author}
  {\bibfnamefont {H.-S.}\ \bibnamefont {Zong}},\ }\href {\doibase
  10.1103/PhysRevLett.115.171801} {\bibfield  {journal} {\bibinfo  {journal}
  {Phys. Rev. Lett.}\ }\textbf {\bibinfo {volume} {115}},\ \bibinfo {pages}
  {171801} (\bibinfo {year} {2015})},\ \Eprint
  {http://arxiv.org/abs/1504.04386} {arXiv:1504.04386 [nucl-th]} \BibitemShut
  {NoStop}%
\bibitem [{\citenamefont {Carrillo-Serrano}\ \emph {et~al.}(2016)\citenamefont
  {Carrillo-Serrano}, \citenamefont {Bentz}, \citenamefont {Cloët},\ and\
  \citenamefont {Thomas}}]{Carrillo-Serrano:2016igi}%
  \BibitemOpen
  \bibfield  {author} {\bibinfo {author} {\bibfnamefont {M.~E.}\ \bibnamefont
  {Carrillo-Serrano}}, \bibinfo {author} {\bibfnamefont {W.}~\bibnamefont
  {Bentz}}, \bibinfo {author} {\bibfnamefont {I.~C.}\ \bibnamefont {Cloët}}, \
  and\ \bibinfo {author} {\bibfnamefont {A.~W.}\ \bibnamefont {Thomas}},\
  }\href {\doibase 10.1016/j.physletb.2016.05.065} {\bibfield  {journal}
  {\bibinfo  {journal} {Phys. Lett.}\ }\textbf {\bibinfo {volume} {B759}},\
  \bibinfo {pages} {178} (\bibinfo {year} {2016})},\ \Eprint
  {http://arxiv.org/abs/1603.02741} {arXiv:1603.02741 [nucl-th]} \BibitemShut
  {NoStop}%
\bibitem [{\citenamefont {Berger}\ \emph {et~al.}(2001)\citenamefont {Berger},
  \citenamefont {Cano}, \citenamefont {Diehl},\ and\ \citenamefont
  {Pire}}]{Berger:2001zb}%
  \BibitemOpen
  \bibfield  {author} {\bibinfo {author} {\bibfnamefont {E.~R.}\ \bibnamefont
  {Berger}}, \bibinfo {author} {\bibfnamefont {F.}~\bibnamefont {Cano}},
  \bibinfo {author} {\bibfnamefont {M.}~\bibnamefont {Diehl}}, \ and\ \bibinfo
  {author} {\bibfnamefont {B.}~\bibnamefont {Pire}},\ }\href {\doibase
  10.1103/PhysRevLett.87.142302} {\bibfield  {journal} {\bibinfo  {journal}
  {Phys. Rev. Lett.}\ }\textbf {\bibinfo {volume} {87}},\ \bibinfo {pages}
  {142302} (\bibinfo {year} {2001})},\ \Eprint
  {http://arxiv.org/abs/hep-ph/0106192} {arXiv:hep-ph/0106192 [hep-ph]}
  \BibitemShut {NoStop}%
\bibitem [{\citenamefont {Shi}\ \emph {et~al.}()\citenamefont {Shi},
  \citenamefont {Bednar}, \citenamefont {Cloët}, \citenamefont {Freese},
  \citenamefont {Mezrag},\ and\ \citenamefont {Zong}}]{shi:inpress}%
  \BibitemOpen
  \bibfield  {author} {\bibinfo {author} {\bibfnamefont {C.}~\bibnamefont
  {Shi}}, \bibinfo {author} {\bibfnamefont {K.}~\bibnamefont {Bednar}},
  \bibinfo {author} {\bibfnamefont {I.~C.}\ \bibnamefont {Cloët}}, \bibinfo
  {author} {\bibfnamefont {A.}~\bibnamefont {Freese}}, \bibinfo {author}
  {\bibfnamefont {C.}~\bibnamefont {Mezrag}}, \ and\ \bibinfo {author}
  {\bibfnamefont {H.}~\bibnamefont {Zong}},\ }\href@noop {} {\ }\BibitemShut
  {NoStop}%
\bibitem [{\citenamefont {Diehl}\ \emph {et~al.}(2001)\citenamefont {Diehl},
  \citenamefont {Feldmann}, \citenamefont {Jakob},\ and\ \citenamefont
  {Kroll}}]{Diehl:2000xz}%
  \BibitemOpen
  \bibfield  {author} {\bibinfo {author} {\bibfnamefont {M.}~\bibnamefont
  {Diehl}}, \bibinfo {author} {\bibfnamefont {T.}~\bibnamefont {Feldmann}},
  \bibinfo {author} {\bibfnamefont {R.}~\bibnamefont {Jakob}}, \ and\ \bibinfo
  {author} {\bibfnamefont {P.}~\bibnamefont {Kroll}},\ }\href {\doibase
  10.1016/S0550-3213(00)00684-2, 10.1016/S0550-3213(01)00183-3} {\bibfield
  {journal} {\bibinfo  {journal} {Nucl. Phys.}\ }\textbf {\bibinfo {volume}
  {B596}},\ \bibinfo {pages} {33} (\bibinfo {year} {2001})},\ \bibinfo {note}
  {[Erratum: Nucl. Phys.B605,647(2001)]},\ \Eprint
  {http://arxiv.org/abs/hep-ph/0009255} {arXiv:hep-ph/0009255 [hep-ph]}
  \BibitemShut {NoStop}%
\bibitem [{\citenamefont {de~Melo}\ \emph {et~al.}(1998)\citenamefont
  {de~Melo}, \citenamefont {Sales}, \citenamefont {Frederico},\ and\
  \citenamefont {Sauer}}]{deMelo:1998an}%
  \BibitemOpen
  \bibfield  {author} {\bibinfo {author} {\bibfnamefont {J.~P. B.~C.}\
  \bibnamefont {de~Melo}}, \bibinfo {author} {\bibfnamefont {J.~H.~O.}\
  \bibnamefont {Sales}}, \bibinfo {author} {\bibfnamefont {T.}~\bibnamefont
  {Frederico}}, \ and\ \bibinfo {author} {\bibfnamefont {P.~U.}\ \bibnamefont
  {Sauer}},\ }\href {\doibase 10.1016/S0375-9474(98)00070-0} {\bibfield
  {journal} {\bibinfo  {journal} {Nucl. Phys.}\ }\textbf {\bibinfo {volume}
  {A631}},\ \bibinfo {pages} {574C} (\bibinfo {year} {1998})},\ \Eprint
  {http://arxiv.org/abs/hep-ph/9802325} {arXiv:hep-ph/9802325 [hep-ph]}
  \BibitemShut {NoStop}%
\bibitem [{\citenamefont {Li}\ \emph {et~al.}(2018)\citenamefont {Li},
  \citenamefont {Maris},\ and\ \citenamefont {Vary}}]{Li:2017uug}%
  \BibitemOpen
  \bibfield  {author} {\bibinfo {author} {\bibfnamefont {Y.}~\bibnamefont
  {Li}}, \bibinfo {author} {\bibfnamefont {P.}~\bibnamefont {Maris}}, \ and\
  \bibinfo {author} {\bibfnamefont {J.}~\bibnamefont {Vary}},\ }\href {\doibase
  10.1103/PhysRevD.97.054034} {\bibfield  {journal} {\bibinfo  {journal} {Phys.
  Rev.}\ }\textbf {\bibinfo {volume} {D97}},\ \bibinfo {pages} {054034}
  (\bibinfo {year} {2018})},\ \Eprint {http://arxiv.org/abs/1712.03467}
  {arXiv:1712.03467 [hep-ph]} \BibitemShut {NoStop}%
\bibitem [{\citenamefont {Petrov}\ \emph {et~al.}(1998)\citenamefont {Petrov},
  \citenamefont {Pobylitsa}, \citenamefont {Polyakov}, \citenamefont {Bornig},
  \citenamefont {Goeke},\ and\ \citenamefont {Weiss}}]{Petrov:1998kf}%
  \BibitemOpen
  \bibfield  {author} {\bibinfo {author} {\bibfnamefont {V.~{\relax Yu}.}\
  \bibnamefont {Petrov}}, \bibinfo {author} {\bibfnamefont {P.~V.}\
  \bibnamefont {Pobylitsa}}, \bibinfo {author} {\bibfnamefont {M.~V.}\
  \bibnamefont {Polyakov}}, \bibinfo {author} {\bibfnamefont {I.}~\bibnamefont
  {Bornig}}, \bibinfo {author} {\bibfnamefont {K.}~\bibnamefont {Goeke}}, \
  and\ \bibinfo {author} {\bibfnamefont {C.}~\bibnamefont {Weiss}},\ }\href
  {\doibase 10.1103/PhysRevD.57.4325} {\bibfield  {journal} {\bibinfo
  {journal} {Phys. Rev.}\ }\textbf {\bibinfo {volume} {D57}},\ \bibinfo {pages}
  {4325} (\bibinfo {year} {1998})},\ \Eprint
  {http://arxiv.org/abs/hep-ph/9710270} {arXiv:hep-ph/9710270 [hep-ph]}
  \BibitemShut {NoStop}%
\bibitem [{\citenamefont {Polyakov}\ and\ \citenamefont
  {Weiss}(1999)}]{Polyakov:1999gs}%
  \BibitemOpen
  \bibfield  {author} {\bibinfo {author} {\bibfnamefont {M.~V.}\ \bibnamefont
  {Polyakov}}\ and\ \bibinfo {author} {\bibfnamefont {C.}~\bibnamefont
  {Weiss}},\ }\href {\doibase 10.1103/PhysRevD.60.114017} {\bibfield  {journal}
  {\bibinfo  {journal} {Phys. Rev.}\ }\textbf {\bibinfo {volume} {D60}},\
  \bibinfo {pages} {114017} (\bibinfo {year} {1999})},\ \Eprint
  {http://arxiv.org/abs/hep-ph/9902451} {arXiv:hep-ph/9902451 [hep-ph]}
  \BibitemShut {NoStop}%
\bibitem [{\citenamefont {Theussl}\ \emph {et~al.}(2004)\citenamefont
  {Theussl}, \citenamefont {Noguera},\ and\ \citenamefont
  {Vento}}]{Theussl:2002xp}%
  \BibitemOpen
  \bibfield  {author} {\bibinfo {author} {\bibfnamefont {L.}~\bibnamefont
  {Theussl}}, \bibinfo {author} {\bibfnamefont {S.}~\bibnamefont {Noguera}}, \
  and\ \bibinfo {author} {\bibfnamefont {V.}~\bibnamefont {Vento}},\ }\href
  {\doibase 10.1140/epja/i2003-10174-3} {\bibfield  {journal} {\bibinfo
  {journal} {Eur. Phys. J.}\ }\textbf {\bibinfo {volume} {A20}},\ \bibinfo
  {pages} {483} (\bibinfo {year} {2004})},\ \Eprint
  {http://arxiv.org/abs/nucl-th/0211036} {arXiv:nucl-th/0211036 [nucl-th]}
  \BibitemShut {NoStop}%
\bibitem [{\citenamefont {Ji}(1997{\natexlab{b}})}]{Ji:1996nm}%
  \BibitemOpen
  \bibfield  {author} {\bibinfo {author} {\bibfnamefont {X.-D.}\ \bibnamefont
  {Ji}},\ }\href {\doibase 10.1103/PhysRevD.55.7114} {\bibfield  {journal}
  {\bibinfo  {journal} {Phys. Rev.}\ }\textbf {\bibinfo {volume} {D55}},\
  \bibinfo {pages} {7114} (\bibinfo {year} {1997}{\natexlab{b}})},\ \Eprint
  {http://arxiv.org/abs/hep-ph/9609381} {arXiv:hep-ph/9609381 [hep-ph]}
  \BibitemShut {NoStop}%
\bibitem [{\citenamefont {Ji}(1998)}]{Ji:1998pc}%
  \BibitemOpen
  \bibfield  {author} {\bibinfo {author} {\bibfnamefont {X.-D.}\ \bibnamefont
  {Ji}},\ }\href {\doibase 10.1088/0954-3899/24/7/002} {\bibfield  {journal}
  {\bibinfo  {journal} {J. Phys.}\ }\textbf {\bibinfo {volume} {G24}},\
  \bibinfo {pages} {1181} (\bibinfo {year} {1998})},\ \Eprint
  {http://arxiv.org/abs/hep-ph/9807358} {arXiv:hep-ph/9807358 [hep-ph]}
  \BibitemShut {NoStop}%
\bibitem [{\citenamefont {Diehl}(2003)}]{Diehl:2003ny}%
  \BibitemOpen
  \bibfield  {author} {\bibinfo {author} {\bibfnamefont {M.}~\bibnamefont
  {Diehl}},\ }\href {\doibase 10.1016/j.physrep.2003.08.002,
  10.3204/DESY-THESIS-2003-018} {\bibfield  {journal} {\bibinfo  {journal}
  {Phys. Rept.}\ }\textbf {\bibinfo {volume} {388}},\ \bibinfo {pages} {41}
  (\bibinfo {year} {2003})},\ \Eprint {http://arxiv.org/abs/hep-ph/0307382}
  {arXiv:hep-ph/0307382 [hep-ph]} \BibitemShut {NoStop}%
\bibitem [{\citenamefont {Burkert}\ \emph {et~al.}(2018)\citenamefont
  {Burkert}, \citenamefont {Elouadrhiri},\ and\ \citenamefont
  {Girod}}]{Burkert:2018bqq}%
  \BibitemOpen
  \bibfield  {author} {\bibinfo {author} {\bibfnamefont {V.~D.}\ \bibnamefont
  {Burkert}}, \bibinfo {author} {\bibfnamefont {L.}~\bibnamefont
  {Elouadrhiri}}, \ and\ \bibinfo {author} {\bibfnamefont {F.~X.}\ \bibnamefont
  {Girod}},\ }\href {\doibase 10.1038/s41586-018-0060-z} {\bibfield  {journal}
  {\bibinfo  {journal} {Nature}\ }\textbf {\bibinfo {volume} {557}},\ \bibinfo
  {pages} {396} (\bibinfo {year} {2018})}\BibitemShut {NoStop}%
\bibitem [{\citenamefont {Kumerički}(2019)}]{Kumericki:2019ddg}%
  \BibitemOpen
  \bibfield  {author} {\bibinfo {author} {\bibfnamefont {K.}~\bibnamefont
  {Kumerički}},\ }\href {\doibase 10.1038/s41586-019-1211-6} {\bibfield
  {journal} {\bibinfo  {journal} {Nature}\ }\textbf {\bibinfo {volume} {570}},\
  \bibinfo {pages} {E1} (\bibinfo {year} {2019})}\BibitemShut {NoStop}%
\end{thebibliography}%

\end{document}